\documentstyle[12pt, amsfonts, amssymb]{article}
\begin{document}


\def\theequation{\thesection.\arabic{equation}}
\def\be{\begin{equation}}
\def\ee{\end{equation}}
\def\ba{\begin{eqnarray}}
\def\ea{\end{eqnarray}}
\def\lb{\label}
\def\nn{\nonumber}

\def\a{\alpha}
\def\b{\beta}
\def\g{\gamma}
\def\d{\delta}
\def\e{\varepsilon}
\def\l{\lambda}
\def\s{\sigma}
\def\L{\Lambda}
\def\E{{\cal E}}
\def\Vp{{\cal V}_p}
\def\Hp{{\cal H}_p}
\def\bq{\overline{q}}

\def\id{\mbox{\rm 1\hspace{-3.5pt}I}}
\newcommand{\ID}[2]{\id^{| #1 {\cal i}}_{\;\;\; {\cal h} #2 |}}

\def\p{\hat p}
\def\C{\Bbb C}
\def\Z{\Bbb Z}
\def\F{\Bbb F}
\def\1{1\!\!{\rm I}}
\def\eod{\phantom{a}\hfill \rule{2.5mm}{2.5mm}}

\def\hF{\hat{F}}
\def\hA{\hat{A}}
\def\hB{\hat{B}}
\newcommand{\dy}[1]{DYBE(${#1}$)}
\newcommand{\xR}[1]{\ ^{\small #1}\!\!\hR}
\newcommand{\xF}[1]{\ ^{\small #1}\!\!\hF}

\newcommand{\rank}{\mathop{\rm rank}\nolimits}
\newcommand{\height}{\mathop{\rm height}\nolimits}
\newcommand{\aut}{\mathop{\rm Aut}\nolimits}
\newcommand{\rx}{\mathop{\rho_{\hspace{-1pt}\scriptscriptstyle W,k}}\nolimits}
\newcommand{\rn}{\mathop{\rho_{\hspace{-1pt}\scriptscriptstyle W,n}}\nolimits}
\newcommand{\q}[1]{[#1]}
\newcommand{\ai}[1]{{a_{#1}}\, }
\newcommand{\ainv}[2]{({a^{-1})}^{| #1 {\cal i}}_{\;\;\; {\cal h} #2 |}}

\def\R{\hat{R}}
\def\Rp{\hat{R}(p)}
\def\p{\hat p}
\newcommand{\DR}[1]{\hat{R}_{#1}(p)}
\newcommand{\DDR}[2]{\hat{R}^{#1}_{#2}(p)}
\newcommand{\DDDR}[2]{\hat{R}^{#1}_{#2}(p')}

\def\A{{A}}

\def\eup{\varepsilon^{|1  \dots n{\cal i}}}
\def\edo{\varepsilon_{{\cal h}1  \dots n|}}
\def\eupp{\varepsilon^{|1  \dots n{\cal i}}(p)}
\def\edop{\varepsilon_{{\cal h}1  \dots n|}(p)}

\def\eu2{\varepsilon^{|2  \dots n{+}1{\cal i}}}
\def\ed2{\varepsilon_{{\cal h}2  \dots n{+}1|}}
\def\eu2p{\varepsilon^{|2  \dots n{+}1{\cal i}}(p)}
\def\ed2p{\varepsilon_{{\cal h}2  \dots n{+}1|}(p)}

\def\Eup{ {\cal E}^{|1  \dots n{\cal i} } }
\def\Edo{ {\cal E}_{{\cal h}1  \dots n| } }
\def\Eupp{{\cal E}^{|1  \dots n{\cal i}}(p)}
\def\Edop{{\cal E}_{{\cal h}1  \dots n|}(p)}

\def\U2{ {\cal E}^{|2  \dots n{+}1{\cal i} } }
\def\D2{ {\cal E}_{{\cal h}2  \dots n{+}1| } }
\def\Eu2p{{\cal E}^{|2  \dots n{+}1{\cal i}}(p)}
\def\Ed2p{{\cal E}_{{\cal h}2  \dots n{+}1|}(p)}

\newcommand{\eupi}[1]{\varepsilon^{|#1 \dots n+ #1 -1{\cal i}}}
\newcommand{\edoi}[1]{\varepsilon_{{\cal h}#1 \dots n+ #1 -1|}}
\newcommand{\Eupi}[1]{{\cal E}^{|#1 \dots n+ #1 -1{\cal i}}(p)}
\newcommand{\Edoi}[1]{{\cal E}_{{\cal h}#1 \dots n+ #1 -1|}(p)}
\newcommand{\N}[2]{N^{| #1 {\cal i}}_{\;\;\; {\cal h} #2 |}}
\newcommand{\K}[2]{K^{| #1 {\cal i}}_{\;\;\;\;\;\;\;\: {\cal h} #2 |}}
\newcommand{\iK}[2]{{K^{-1}}^{| #1 {\cal i}}_{\;\;\; {\cal h} #2 |}}

\def\bbr{{\rm I}\!{\rm R}}
\def\bbz{{\rm Z}\!\!\!{\rm Z}}
\def\subbbc{{\rm C}\kern-3.3pt\hbox{\vrule height4.8pt width0.4pt}\,}
\def\BbbZ{Z\!\!\! Z}
\def\BbbN{{\rm I}\!{\rm N}}

\begin{center}


{\Huge\bf Quantum matrix algebra}\\[3 mm]
{\Huge\bf for the $SU(n)$ WZNW model}\\[18 mm]


{\large{\bf
P. Furlan$^{a,b}$
\footnote[1]{e-mail address: furlan@trieste.infn.it},
L.K. Hadjiivanov$^{c,b}$
\footnote[2]{e-mail address: lhadji@inrne.bas.bg},
A.P.~Isaev$^d$
\footnote[3]{e-mail address: isaevap@thsun1.jinr.ru},\\
O.V. Ogievetsky$^{e}$
\footnote[4]{On leave of absence from P.N. Lebedev Physical Institute, Theoretical
department, 117924 Moscow, Leninsky prospect 53, Russia; e-mail address:
oleg@cpt.univ-mrs.fr},
P.N. Pyatov$^d$
\footnote[5]{e-mail address: pyatov@thsun1.jinr.ru},
I.T. Todorov$^{c}$
\footnote[6]{e-mail address: todorov@inrne.bas.bg}\\}}
\vskip 0.5 cm
$^a$Dipartimento di Fisica
Teorica dell' Universit\`a degli Studi\\ di
Trieste, Strada Costiera 11, I-34014 Trieste, Italy\\
$^b$Istituto Nazionale di Fisica Nucleare (INFN),
Sezione di Trieste,\\
Trieste, Italy\\
$^c$Theoretical Physics Division, Institute for Nuclear
Research and\\ Nuclear Energy, Tsarigradsko Chaussee
72, BG-1784 Sofia, Bulgaria\\
$^d$Bogoliubov Laboratory of Theoretical
Physics, JINR, Dubna,\\
141 980 Moscow Region, Russia\\
$^e$Centre de Physique Th\'eorique, Luminy, F-13288 Marseille, France

\end{center}


\vspace{1cm}

\noindent
{\bf PACS:} 02.10.Tq, 11.10Kk, 13.30.Ly

\noindent
{\bf Keywords:}

\noindent
{WZNW model, quantum matrix algebra,
quantum gauge group, model space}

\vspace{5mm}

\noindent
{\Large{\bf Running title:}}

\noindent
{\Large{Quantum matrix algebra for the $SU(n)$ WZNW model}}

\vspace{1cm}

\newpage


\begin{abstract}

{\normalsize
\noindent
The zero modes of the chiral $SU(n)$ WZNW model give rise to
an {\em intertwining quantum matrix algebra}
$\cal A$ generated by an $n\times n$ matrix $a=\left( a^i_\alpha
\right)\,,i, \a =1,\dots ,n\,$ (with noncommuting entries) and by
rational functions of $n$ commuting elements $q^{p_i}$ satisfying
$\prod\limits_{i=1}^n q^{p_i} = 1,\,\, q^{p_i} a^j_\a = a^j_\a
q^{p_i + \delta^j_i - {1\over n}}\,.$ We study a generalization of
the Fock space (${\cal{F}}$) representation of ${\cal A}\,$ for
generic $q\,$ ($q$ not a root of unity) and demonstrate that it
gives rise to a {\em model} of the quantum universal enveloping
algebra $U_q = U_q(sl_n)\,$ each irreducible representation
entering ${\cal{F}}$ with multiplicity $1\,$. For an integer
${\widehat{su}}(n)$ {\em height} $h (=k+n\ge n)$ the complex
parameter $q$ is an even root of unity, $q^h=-1\,,$ and the algebra
${\cal A}\,$ has an ideal ${\cal{I}}_h\,$ such that the factor
algebra ${\cal A}_h ={\cal A}/{\cal I}_h$ is finite dimensional.
All physical $U_q\,$ modules -- of shifted weights satisfying
$p_{1n}\equiv p_1 - p_n < h\,$ -- appear in the Fock representation
of ${\cal A}_h\;.$}

\end{abstract}

\newpage


\textwidth = 16truecm
\textheight = 22truecm
\hoffset = -1truecm
\voffset = -2truecm

\section*{Introduction}
\setcounter{equation}{0}
\renewcommand\theequation{0.\arabic{equation}}

\medskip

Although the Wess-Zumino-Novikov-Witten
(WZNW) model was first formulated in terms of a (multivalued)
action \cite{W}, it was originally solved \cite{KZ}
by using axiomatic conformal field theory methods.
The two dimensional (2D) Euclidean Green functions have been expressed
\cite{BPZ} as sums of products of analytic and antianalytic {\em
conformal blocks}. Their operator interpretation exhibits some
puzzling features: the presence of {\em noninteger} ("quantum")
{\em statistical dimensions} (that appear as positive real
solutions of the {\em fusion rules} \cite{V}) contrasted with the
local ("Bose") commutation relations (CR) of the corresponding 2D
fields. The gradual understanding of both the factorization
property and the hidden {\em braid group statistics} (signaled by
the quantum dimensions) only begins with the development of the
canonical approach to the model (for a sample of references, see
\cite{B, F1, F2, G, FG, Chu, FHT1, FHT2, FHT3, FHT6}) and
the associated splitting of the basic group
valued field $g : \; {\Bbb S}^1 \times {\Bbb R}\;\rightarrow\;
G\;$ into chiral parts. The resulting zero mode extended phase
space displays a new type of {\em quantum} group {\em gauge
symmetry}: on one hand, it is expressed in terms of the {\em
quantum universal enveloping algebra} $U_q({\cal G})\,,$ a
deformation of the {\em finite dimensional Lie algebra} ${\cal
G}\,$ of $G\,$ -- much like a gauge symmetry of the first kind; on
the other, it requires the introduction of an {\em extended,
indefinite metric state space}, a typical feature of a (local)
gauge theory of the second kind.

Chiral fields admit an expansion into {\em chiral vertex operators}
(CVO) \cite{TK} which
diagonalize the monodromy and are expressed in terms of
the currents' degrees
of freedom with "zero mode" coefficients that are independent of
the world sheet coordinate \cite{AF, BF, FHT2, FHT3, FHT6}. Such a type
of quantum theory has been studied in the framework of lattice
current algebras (see \cite{F2, G, FG, AFFS, BS} and references
therein). Its accurate formulation in the continuum limit has only
been attempted in the case of $G=SU(2)$ (see \cite{FHT2, FHT3,
DT}). The identification (in \cite{HIOPT}) of the zero mode
($U_q$)vertex operators $a^i_\a\,$ (the "$U_q$-oscillators" of the
$SU(2)\,$ case \cite{FHT2}) with the generators of a {\em quantum
matrix algebra} defined by a pair of (dynamical) $R$-matrices
allows to extend this approach to the case of $G = SU(n)\,$.

The basic group valued chiral field $u^A_\a (x)$ is thus expanded
in CVO $u^A_i (x,p)$ which interpolate between chiral current
algebra modules of weight $p=p_j v^{(j)}$ and $p+v^{(i)}\,,\
i=1,\dots ,n\,$ (in the notation of \cite{HIOPT} to be
recapitulated in Section 1 below). The operator valued coefficients
$a^i_\a$ of the resulting expansion intertwine finite dimensional
irreducible representations (IR) of $U_q \equiv U_q(sl_n)\,$ that
are labeled by the same weights. For generic $q\,$ ($q\,$ not a
root of unity) they generate, acting on a suitably defined vacuum
vector, a Fock-like space ${\cal F}\,$ that contains every (finite
dimensional) IR of $U_q\,$ with multiplicity $1 ,\,$ thus providing
a {\em model} for $U_q\,$ in the sense of \cite{BGG}. This result
(established in Section 3.1) appears to be novel even in the
undeformed case ($q = 1$) giving rise to a new (for $n > 2$) model
of $SU(n)\,.$ In the important case of $q\,$ an even root of unity
($q^h =-1\,$) we have prepared the ground (in Sections 3.2 and 3.3)
for a (co)homological study of the two dimensional (left and right
movers') zero mode problem \cite{DT}.

It should be emphasized that displaying the quantum group's degrees of
freedom requires an extension of the phase space of the models under 
consideration. Much interesting work on both physical and mathematical
aspects of $2D\,$ conformal field theory has been performed without going
to such an extension -- see e.g. \cite{BPZ, KZ, MS, BK, FBZ}. The concept 
of a quantum group, on the other hand, has emerged in the study of closely 
related integrable systems and its uncovering in conformal field theory models 
has fascinated researchers from the outset -- see e.g. \cite{B, BBB, PS, F1}. (For a
historical survey of an early stage of this development see \cite{GS}. Significant
later developments in different directions -- beyond the scope of the present paper 
-- can be found e.g. in \cite{FK, MaS, BNS, PZ}.)

Even within the scope of this paper there remain unresolved problems.
We have, for instance, no operator realization of the extended chiral WZNW 
model, involving indecomposable highest weight modules of the Kac-Moody
current algebra.

The paper is organized as follows. Section 1 provides an updated
summary of recent work \cite{FHT1} - \cite{FHT3} on the $SU(n)$
WZNW model. A new point here is the accurate treatment of the path
dependence of the exchange relations in both the $x\,$ and the $z=
e^{ix}\,$ pictures (Proposition 1.3). In Section 2 we carry out the
factorization of the chiral field $u(x)$ into CVO and $U_q$ vertex
operators and review relevant results of \cite{HIOPT} computing, in
particular, the determinant of the quantum matrix $a$ as a function
of the $U_q(sl_n)$ weights. The discussion of the interrelation
between the braiding properties of four-point blocks and the
exchange relations among zero modes presented in Section 2.2 is
new; so are some technical results like Proposition 2.3 used in the
sequel. Section 3.1 introduces the Fock space (${\cal F}$)
representation of the zero mode algebra ${\cal A}$ for generic
$q\;$; the main result is summed up in Proposition 3.3. In Section
3.2 we compute inner products for the canonical bases in the
$U_q\,$ modules ${\cal F}_p\,$ for $n=2,3\,.$ In Section 3.3 we
study the kernel of the inner product in ${\cal{F}}$ for $q$ an
even root of unity,
\be
\lb{0.1}
q=e^{-i{\pi\over h}} \quad (h=k+n\ge n)\, .
\ee
It is presented in the form ${\tilde{\cal{I}}}_h{\cal{F}}$ where
${\tilde{\cal{I}}}_h$ is an ideal in ${\cal A}\,.$ We select a
smaller ideal ${\cal{I}}_h \subset {\tilde{\cal{I}}}_h\,$
(introduced in \cite{HIOPT}) such that the factor algebra ${\cal
A}_h = {\cal A}/{\cal I}_h\,$ is still finite-dimensional but
contains along with each physical weight $p\,$ (with $p_{1 n}<
h\,$) a weight ${\tilde p}\,$ corresponding to the first singular
vector of the associated Kac-Moody module (cf. Remark 2.1).

\vspace{5mm}


\section{Monodromy extended $\; SU(n)$ WZNW model: a synopsis}
\subsection{Exchange relations; path dependent monomials of
chiral fields}

\medskip

\setcounter{equation}{0}
\renewcommand{\theequation}{\thesection.1\alph{equation}}

The WZNW action for a group valued field on a cylindric space-time
${\Bbb R}^1\times {\Bbb S}^1\,$ is written as
\be
\lb{1.1a}
S=- {k\over{4\pi}} \int\{{\rm tr} (g^{-1}\partial_+ g)
(g^{-1}\partial_- g) dx^+ dx^- + s^*\omega (g)\}\,,\quad x^\pm =
x\pm t
\ee
where $s^*\omega\,$ is the pullback
($s^* g^{-1} dg = g^{-1} \partial_+ g dx^+ + g^{-1} \partial_- g dx^-\,$) of a
two-form $\omega\;$ on $G\;$ satisfying
\be
\lb{1.1b}
d\omega (g) = {1\over 3} {\rm tr} (g^{-1} dg)^3\,.
\ee

The general, $G=SU(n)$ valued (periodic) solution,
$\; g(t,x+2\pi)=g(t,x),$ of the resulting equations of motion
factorizes into a product of group valued chiral fields
\setcounter{equation}{0}
\renewcommand{\theequation}{\thesection.2\alph{equation}}
\be
\lb{1.1}
g^A_B(t,x)=u^A_\alpha (x+t)({\bar u}^{-1})^\alpha_B (x-t)\quad
({\rm classically,}\ g, u, {\bar u}\in SU(n)),
\ee
where $u$ and $\bar u$ satisfy a twisted periodicity condition
\be
\lb{1.2}
u(x+2\pi )=u(x)M\,,\,\,\,{\bar u}(x+2\pi )={\bar u}(x){\bar M}
\ee
with {\em equal monodromies}, $\bar M = M\,.$
\setcounter{equation}{2}
\renewcommand{\theequation}{\thesection.\arabic{equation}}
The symplectic form of the $2D\,$ model is
expressed as a sum of two chiral $2$-forms involving the monodromy:
\be
\lb{1.3}
\Omega^{(2)} = \Omega (u,M) - \Omega ({\bar u}, M)\,,
\ee
$$
\Omega (u,M) = {k\over 4\pi}\ \left( {\rm tr}\left(\,
\int\limits^\pi_{-\pi}
\partial (u^{-1}\, du) u^{-1}\ du\ dx
- b^{-1} db dM M^{-1}\right) + \rho (M)\right)\,.
$$
Here $b = u(-\pi)\,$ and the $2$-form $\rho (M)$ is restricted by the
requirement that
$\Omega (u,M)\,$ is closed,
$d\;\Omega (u,M)\,= \, 0$ which is equivalent to
\setcounter{equation}{0}
\renewcommand{\theequation}{\thesection.4\alph{equation}}
\be
d \rho (M) = {1\over 3}\; {\rm tr}\; (dMM^{-1})^3
\ee
(in other words, $\rho\;$ satisfies the same equation (1.1b) as
$\omega\;$).

Such a $\rho\,$ can
only be defined locally -- in an open
dense neighbourhood of the identity of the complexification
of $SU(n)\,$ to $SL(n,\C )\,$.
An example is given by
\be
\rho (M)\, = \,{\rm tr}\, (M_+^{-1} dM_+ M_-^{-1} dM_- )
\ee
where $M_\pm\,$ are the Gauss components
of $M$
(which are well defined for
$M_{nn}\ne 0 \ne {\rm det} \left(\matrix{M_{n-1n-1}& M_{n-1n}\cr
M_{nn-1}&M_{nn}\cr}\right)$ etc.):
\setcounter{equation}{4}
\renewcommand{\theequation}{\thesection.\arabic{equation}}
\be
\label{1.5}
M=M_+M_-^{-1}\,,\quad
M_+ = N_+ D\,,\quad M_-^{-1} = N_- D\,,
\ee

\be
\lb{1.6}
N_+= \left(\matrix{ 1 &f_1 &f_{12} &\ldots\cr
0 &1 &f_2 &\ldots\cr
0 &0 &1 &\ldots\cr
\ldots &\ldots &\ldots &\ldots
\cr}\right),\,
N_-= \left(\matrix{ 1 &0 &0 &\ldots\cr
e_1 &1 &0 &\ldots\cr
e_{21} &e_2 &1 &\ldots\cr
\ldots &\ldots &\ldots &\ldots\cr}\right)\,,
D=(d_{\alpha}{\delta}^{\alpha}_{\beta})\,,\nonumber
\ee
and the common diagonal matrix $D$ has unit determinant:
$d_1d_2\ldots d_n=1\,.$ Different solutions $\rho$ of (1.4a) correspond to
different non-degenerate solutions of the classical
Yang-Baxter equation \cite{G, FHT1}.

The closed $2$-form (\ref{1.3}) on the space of chiral variables $u, \bar
u , M\,$ is degenerate.  This fact is related to the non-uniqueness of the
decomposition (\ref{1.1}): $g(t,x)\,$ does not change under constant right
shifts of the chiral components, $ u\ \to \ u h\,,\ {\bar u}\ \to \ {\bar
u} h\,$, $h\in G$. Under such shifts the monodromy changes as
$\ M \ \to \ h^{-1} M h\,$ (see also the discussion of
this point in \cite{BFP}).  We restore non-degeneracy by further extending
the phase space, assuming that the monodromies $M\,$ and $\bar M\,$ of
$u\,$ and $\bar u\,$ are independent so that the left and the right sector
completely decouple. As a result monodromy invariance in the extended
phase space is lost since $M\,$ and $\bar M\,$ satisfy Poisson bracket
relations of opposite sign (due to (\ref{1.3})) and hence cannot be
identified. Singlevaluedness of $g(t,x)\,$ can only be recovered in a
weak sense, when $g\,$ is applied to a suitable subspace of "physical
states" in the quantum theory \cite{FHT2, FHT3, DT, goslar}.

We require that
quantization respects all
symmetries of the classical chiral theory.
Apart from conformal invariance and invariance under
periodic left shifts the $(u,M)$ system
admits a Poisson--Lie symmetry under constant right shifts
\cite{S-T-S, AT, FHT1}
which gives rise to a quantum group symmetry in the
quantized theory. The quantum exchange relations
so obtained \cite{F1, F2, G, FG, FHT1, FHT2, FHT3},
\be
\lb{uuR1}
u_2(y)u_1(x)=u_1(x)u_2(y)R(x-y)\,,\quad
{\bar u}_1(x){\bar u}_2(y)={\bar u}_2(y){\bar u}_1(x)R(x-y)
\ee
(for $0< |x-y| < 2\pi\,$) can be also written as braid relations:
\setcounter{equation}{0}
\renewcommand{\theequation}{\thesection.7\alph{equation}}
\ba
\lb{uuR2}
&&P \, u_1(y)\, u_2(x) = u_1(x)\, u_2(y)\, \R (x-y)\\
&&{\bar u}_1^{-1}(y)\,{\bar u}_2^{-1}(x) \,P = \R^{-1} (x-y){\bar
u}_1^{-1}(x)\,{\bar u}_2^{-1}(y)\nonumber\\
\ \Leftrightarrow\
&&{\bar u}_1(x)\,{\bar u}_2(y) \,P = {\bar u}_2 (y)\,
{\bar u}_1(x) \R (x-y)\,.
\ea
Here $R (x)$ is related to the (constant, Jimbo) $SL(n)\ R$-matrix
\cite{J} by
\setcounter{equation}{0}
\renewcommand{\theequation}{\thesection.8\alph{equation}}
\ba
\lb{Jimbo}
R(x) &=& R\,\theta (x) + P R^{-1} P\,\theta (-x)\\
R^{\alpha_{1}\alpha_{2}}_{\beta_{1}\beta_{2}} &=&
{\bar q}^{1\over n} \left(
\delta^{\alpha_{1}\alpha_{2}}_{\beta_1\beta_{2}}\;
q^{\delta_{\alpha_{1}\alpha_{2}}} + (q-\bq )\;
\delta^{\alpha_{1}\alpha_{2}}_{\beta_2\beta_{1}}\;
\theta_{\alpha_{1}\alpha_{2}} \right)
\ea
where
\be
\lb{theta}
\delta^{\alpha_{1}\alpha_{2}}_{\beta_1\beta_{2}}\;=
\delta^{\alpha_{1}}_{\beta_{1}}\delta^{\alpha_2}_{\beta_{2}}\;,\quad
\theta_{\alpha \beta} =
\left\{
\begin{array}{ccc}
1 & {\rm if} & \alpha > \beta \\
0 & {\rm if} & \alpha \le \beta
\end{array}
\right .
\,,\quad \bq := q^{-1}\,,
\ee
$P$ stands for permutation of factors in $V \otimes V\;,\ V = {\C}^n\,,$
while $\R$ is the corresponding braid operator:
\setcounter{equation}{0}
\renewcommand{\theequation}{\thesection.9\alph{equation}}
\be
\lb{P1}
\R = R\, P\,,\quad P \left( X_{|1{\cal i}}\, Y_{|2{\cal i}} \right) = X_{|2{\cal i}}
Y_{|1{\cal i}}
\ee
\be
\lb{P2}
\R (x) = R(x) P =
\left\{
\begin{array}{ccc}
\R & {\rm for} & x > 0 \\
{\R}^{-1} & {\rm for} & x < 0
\end{array}
\right .
\,.
\ee

We are using throughout the tensor product notation of Faddeev et
al. \cite{FRT} : $u_1 = u \otimes \1\,,\ \ u_2 = \1 \otimes u$ are
thus defined as operators in $V \otimes V\,.$

Restoring all indices we can write Eq.(1.7a) as
$$
u^B_\a (y)  u^A_\b (x)\, =\, u^A_\sigma (x)  u^B_\tau (y) \R
(x-y)^{\sigma\tau}_{\a\b}\,.
\eqno{(1.7c)}
$$

\setcounter{equation}{9}
\renewcommand{\theequation}{\thesection.\arabic{equation}}

Whenever dealing with a tensor product of $3$ or more copies of
$V$ we shall write $R_{ij}$ to indicate that $R$ acts
non-trivially on the $i$-th and $j$-th factors (and reduces to
the identity operator on all others).

\vspace{5mm}
\noindent
{\em Remark 1.1~} The operator $\R$ (\ref{P1}) coincides with
${\R}_{21} = P {\R}_{12} P\ \ (P = P_{12} )$ in the notation of
\cite{FRT} and \cite{HIOPT}.
We note that if ${\R}_{i i+1}$ satisfy the Artin braid relations
then so do ${\R}_{i+1 i}\,$; we have, in particular,
\be
\lb{RRR}
{\R}_{12}{\R}_{23}{\R}_{12}={\R}_{23}{\R}_{12}{\R}_{23}\ \Leftrightarrow
\ {\R}_{32}{\R}_{21}{\R}_{32}={\R}_{21}{\R}_{32}{\R}_{21}\,.
\ee
Indeed, the two relations are obtained from one another by acting
from left and right on both sides with the permutation operator
$P_{13} = P_{12} P_{23} P_{12} = P_{23} P_{12} P_{23} \ ( = P_{31})$
and taking into account the identities
\be
\lb{PRP}
P_{13} {\R}_{12} P_{13} = {\R}_{32}\,,\quad
P_{13} {\R}_{23} P_{13} = {\R}_{21}\,.
\ee
Here we shall stick, following Refs.
\cite{FHT1, FHT2, FHT3}, to the
form (1.7), (1.9) of the basic exchange relations.
Note however that (1.1a) involves a change of sign in the WZ term
(as compared to \cite{FHT1, FHT2, FHT3}) which yields the exchange of the
$x^+$ and $x^-\;$ factors in (1.2a) and is responsible for the
sign change in the phase of $q\;$ (\ref{0.1}).

\vspace{5mm}

The multivaluedness of chiral fields requires a more precise
formulation of (1.7). To give an unambiguous meaning to such
exchange relations we shall proceed as follows.

Energy positivity implies that for any $l > 0\,$ the vector
valued function
$$\Psi (\zeta_1,\dots, \zeta_l )\,=\, u_1(\zeta_1)\ldots
u_l(\zeta_l)| 0 {\cal i}$$ is (single valued) analytic on a simply
connected open subset
$$\{\zeta_j=x_j+iy_j\,;\
|x_{j}| < \pi\,,\ j=1,\dots ,l\,;\ y_j<y_{j+1}\,,\ j=1,\dots ,l-1
\}\,,
$$
($x_{jk}:=x_j-x_k$) of the manifold ${\C}^l \setminus {\rm Diag}$
where Diag is defined as the partial diagonal set in ${\C}^l\;:\
{\rm Diag} = \{ (\zeta_1 ,\ldots ,\zeta_l )
\,,\ \ \zeta_j =
\zeta_k\ {\rm for\ some}\ j\ne k\}\,.$

Introduce (exploiting reparametrization invariance -- cf. \cite{FSoT})
the {\em analytic ($z$-) picture} fundamental chiral field
\setcounter{equation}{11}
\renewcommand{\theequation}{\thesection.\arabic{equation}}
\be
\lb{analytic}
\varphi (z)=e^{-i\Delta\zeta} u(\zeta )\,, \quad
z=e^{i\zeta}\,,\quad
\Delta = {{n^2 -1}\over{2hn}}\,,
\ee
$\Delta$ standing for
the conformal dimension of $u\,,$ and note that the variables $z_j\,$ are
radially ordered in the domain ${\cal O}_l\,$:
\be
\lb{domain}
{\cal O}_l = \{z_j = e^{-y_j +ix_j};\,
|z_j| >|z_{j+1}|\,,j=1,\dots ,l-1;\,
|{\rm arg} {z_j}| < \pi ,\, j=1,\dots ,l \}.
\ee

\vspace{5mm}
\noindent
{\em Remark 1.2~} The {\em time evolution} law
\setcounter{equation}{0}
\renewcommand{\theequation}{\thesection.14\alph{equation}}
\be
\lb{L}
e^{itL_0}\; u(x)\; e^{-itL_0}\; =\; u(x+t)
\ee
for the {\em "real compact picture" field} $u(x)$ implies
\be
\lb{L-analytic}
e^{itL_0}\; \varphi (z)\;  e^{-itL_0}\;
 =\; e^{it\Delta}\;\varphi (z e^{it})\,.
\ee
Energy positivity, combined with the prefactor in (\ref{analytic}),
guarantees that the state vector $\varphi (z) | 0 {\cal i}\, $ is a
single valued analytic function of $z$ in the neighbourhood of the
origin (in fact, for a suitably defined inner product, its Taylor
expansion around $z=0$ is norm convergent for $|z|<1$ -- see
\cite{DGM}).

\vspace{5mm}

The vector valued functions
\setcounter{equation}{0}
\renewcommand{\theequation}{\thesection.15\alph{equation}}
\be
\lb{1.15a}
\Phi (z_1,\ldots , z_l ) =
\varphi_1(z_1) \ldots \varphi_l(z_l) |0{\cal i}
\ee
and
\be
\lb{1.15b}
\Psi (\zeta_1,\dots, \zeta_l ) =
u_1(\zeta_1)\ldots u_l(\zeta_l)\, |0 {\cal i}=
\prod\limits_j e^{i\Delta\zeta_j}\, \Phi (e^{i\zeta_1},\ldots ,
e^{i\zeta_l})
\ee
are both analytic in their respective domains (cf. (\ref{domain}))
and are real
analytic (and still single valued) on the parts
\setcounter{equation}{0}
\renewcommand{\theequation}{\thesection.16\alph{equation}}
$$\{ \zeta_j = x_j\ (\;\Rightarrow z_j = e^{ix_j}\;) \,,\
x_1 >x_2 >\ldots >x_l\,,\ x_{1l} <\pi\,\}
$$
of their physical boundaries.

The following Proposition allows to continue these
boundary values through
the domain ${\cal O}_l$ to any other ordered set of $x_j\,$
(the result will be a path dependent multivalued function for
$\{z_1,\ldots ,z_l\}\in {\Bbb C}^l\setminus {\rm Diag}$).

\vspace{5mm}
\noindent
{\bf Proposition 1.3~} {\em
Let $z_1 = e^{ix_1}\,,\, z_2 = e^{ix_2}\,,\ 0<x_{12} < 2\pi\,$;
the path exchanging $x_1\,$ and $x_2\,$ (and hence, $z_1\,$ and
$z_2\,$),
\be
\lb{1.16a}
C_{12}\,:\, \zeta_{1,2}(t)=e^{-i{\pi\over 2}t}\, (x_{1,2}\,{\rm
cos}\, {\pi\over 2}t +i x_{2,1}\, {\rm sin}\, {\pi\over 2}t
)\,,\qquad 0\le t\le 1
\ee
turns clockwise around the middle of the segment $(x_1 , x_2)\,:$
\be
\lb{1.16b}
\zeta_1(t) + \zeta_2(t) = x_1+x_2\,,\
\zeta_{12}(t) := \zeta_1(t)-\zeta_2(t) = x_{12} e^{-i\pi t}
\,.
\ee
Furthermore, if $z_a(t) = e^{i\zeta_a(t)}\,,\ a=1,2\,,$ then
\be
\lb{1.16c}
|z_1(t)|^2 = e^{x_{12}\;{\sin}{\pi}t} = |z_2(t)|^{-2}
 > 1\quad{\rm for}\quad 0<t<1
\ee
so that the pair $( z_1(t) , z_2 (t) )\,$ satisfies the
requirement (\ref{domain}) for two consecutive arguments in the
analyticity domain ${\cal O}_l\,.$
For $0< x_{21}< 2\pi\,$ one has to change the sign of $t\,$ (and
thus the orientation of the path (1.16)) in order to preserve the
inequality $|z_1(t)| > |z_2(t)|\,.$
}

\vspace{5mm}

\noindent
{\em Proof~} All assertions are verified by a direct computation; in
particular, (1.16b) implies
\be
\lb{1.16d}
2\; {\rm Im}\;\zeta_2(t) \;=\; x_{12}\;{\rm sin}\;{\pi t}
\;=\; -2\;{\rm Im}\;\zeta_1(t)
\ee
which yields (1.16c).
\eod

\vspace{5mm}

\setcounter{equation}{16}
\renewcommand{\theequation}{\thesection.\arabic{equation}}
We note that for $\zeta_{1,2}\,$ given by (1.16a) one has
$|\zeta_1(t)|^2+|\zeta_2(t)|^2=x_1^2+x_2^2\,.$
Proposition 1.3 supplements (\ref{analytic}) in describing the
relationship (the essential equivalence) between the real compact
and the analytic picture allowing us to use each time the one
better adapted to the problem under consideration.

We are now prepared to give an unambiguous formulation of the
exchange relations (1.7).

Let ${\Pi}_{12}\; u_1(x_2)\; u_2(x_1)\ \
\left(\, {\Pi}_{12}\;\varphi_1(z_2)\;\varphi_2(z_1)\, \right)\,$
be the analytic continuation of $u_1(x_1)\; u_2(x_2)\,$
(respectively, $\varphi_1(z_1)\;\varphi_2(z_2)$ ) along a path in
the homotopy class of $C_{12}\,$ (1.16). Then Eq.(1.7a)
should be substituted by
\newpage
$$
P \; {\Pi}_{12}\; u_1(x_2)\; u_2(x_1)\; =\; u_1(x_1)\; u_2(x_2)\;
\R \,,
$$
$$~\eqno{(1.7d)}$$
$$
P \;{\Pi}_{12}\; \varphi_1(z_2)\; \varphi_2 (z_1) \; =\;
\varphi_1 (z_1)\; \varphi_2 (z_2)\; \R
$$
for $z_j=e^{ix_j}\,,\, 0<x_{12}<2 \pi\,.$ For
$0<x_{21}<2\pi\,$ and a positively oriented path
one should replace $\R\,$ by ${\R}^{-1}\,.$

We recall (see \cite{FHT3}) that the
quantized $u\,$ (and $g\,$) cannot be treated as group elements. We
can just assert that the operator product expansion of $u\,$ with
its conjugate only involves fields of the family (or, rather, the
Verma module) of the unit operator. The relation
\setcounter{equation}{16}
\renewcommand{\theequation}{\thesection.\arabic{equation}}
\be
\lb{1.17}
u(x+2\pi )\, =\, e^{2\pi i L_0}\, u(x)\, e^{-2\pi i L_0} \,
=\, u(x)\, M\,,
\ee
on the other hand, gives (by (1.14) for $\Delta\,$ given by
(\ref{analytic}))
\setcounter{equation}{0}
\renewcommand{\theequation}{\thesection.18\alph{equation}}
\be
\lb{M-a}
\left( M^\a_\b - q^{{1\over n}-n} {\d}^{\a}_{\b}\right)\, |0>\, =\,
0\,;
\ee
hence, in order to preserve the condition $d_1\dots d_n = 1\,$ for the
product of diagonal elements of $M_+\,$ and $M_-^{-1}\,$ we should
substitute (\ref{1.5}) by its quantum version
\be
\lb{M-b}
M\, =\, q^{{1\over n}-n} M_+ M_-^{-1}\,.
\ee

The tensor products of Gauss components, $M_{2\pm}M_{1\pm}$,
of the monodromy matrix
commute with the braid operator,
\setcounter{equation}{0}
\renewcommand{\theequation}{\thesection.19\alph{equation}}
\ba
[\R\,, M_{2\pm} M_{1\pm} ] = &0& = [\R\,, {\bar M}_{1\pm} {\bar
M}_{2\pm} ]\ ,
\ea
(and hence, with its inverse) but
\setcounter{equation}{0}
\renewcommand{\theequation}{\thesection.19b}
\ba
\R\, M_{2-} M_{1+} = M_{2+} M_{1-} \R\,&,& \quad
\R {\bar M}_{1+}{\bar M}_{2-} = {\bar M}_{1-}{\bar M}_{2+}\,\R
\ea
while the exchange relations between $u$ and $M_{\pm}$ can be
written in the form (cf. \cite{FHT1} - \cite{FHT3})
\setcounter{equation}{19}
\renewcommand{\theequation}{\thesection.\arabic{equation}}
\be
\lb{MPu}
M_{1\pm}\, P\, u_1(x) = u_2(x)\, {\R}^{\mp 1}\, M_{2\pm}\,,\quad
{\bar M}_{2\pm} \,P\, {\bar u}_2(x) =
{\bar u}_1(x)\, \R^{\pm}\, {\bar M}_{1\pm}\,.
\ee
The left and right sectors decouple completely as a consequence
of the
separation of variables in
the classical extended phase space,
\be
\lb{leftright}
[M_1\,, {\bar u}_2 ] = [u_1\,, {\bar u}_2 ] = [M_1\,, {\bar M}_2]
= [ u_1\,, {\bar M}_2 ] = 0\,.
\ee
The above relations for the left sector variables $( u , M )$
are invariant under the left coaction of $SL_q(n)\,,$
\setcounter{equation}{0}
\renewcommand{\theequation}{\thesection.22\alph{equation}}
\be
\lb{QGleft}
u^A_\a (x)\rightarrow (T^{-1})^\b_\a\otimes u^A_\b \equiv (u^A(x)
T^{-1})_\a\,,\ \ M^\a_\b\rightarrow T^\a_\gamma (T^{-1})^\delta_\b
\otimes M^\gamma_\delta \equiv (T M T^{-1} )^\a_\b
\ee
while the right sector is invariant under its right coaction,
\be
\lb{QGright}
{\bar u}^\a_A(x)\rightarrow {\bar u}^\b_A(x)\otimes({\bar T})^\a_\b
= ({\bar T}{\bar u}_A(x))^\a\,,\ \ {\bar M}^\a_\b \rightarrow
{\bar M}^\gamma_\delta\otimes{\bar T}^\a_\gamma
({\bar T}^{-1})^\delta_\b\equiv ({\bar T} {\bar M}{\bar T}^{-1})^\a_\b
\ee
provided
\setcounter{equation}{22}
\renewcommand{\theequation}{\thesection.\arabic{equation}}
\be
\lb{RTT}
\R\, T_2\, T_1 = T_2 \,T_1\, \R\,,\quad \R\, {\bar T}_1\, {\bar T}_2
= {\bar T}_1\, {\bar T}_2 \,\R
\ee
where we have used concise notation in the right hand side
of (1.22). The elements $T^{\alpha}{}_{\beta}$ of $T$ commute
with $u$, $M$, ${\bar u}$ and ${\bar M}$.
The fact that the maps (1.22a) and (1.22b) are respectively
left and right coactions \cite{HIOPT} can be proven by checking
the comodule axioms, see e.g. \cite{Hopf, I1}. There are corresponding
transformations of the elements of $M_\pm\,$ and ${\bar M}_\pm\;$.

Thus the Latin and Greek indices of $u$ and ${\bar u}$ in (\ref{1.1})
transform differently: {\footnotesize A, B}
correspond to the (undeformed) $SU(n)$
action while {\small{$\alpha$}} is a quantum group index.

It is known, on the other hand, that
the first equations in (1.19a) and (1.19b) for
the matrices $M_{\pm}$ are equivalent to the defining
relations  of the ("simply connected"
\cite{DCK}) quantum universal enveloping algebra
(QUEA) $U_{q}(sl_n)$ that is paired by duality to
$Fun\;( SL_{q}(n) )$ (see \cite{FRT}). The Chevalley generators of
$U_q\,$ are related to the elements $d_i, \, e_i,\, f_i$
of the matrices (\ref{1.5})
by (\cite{FRT}; see also \cite{FHT3})
\setcounter{equation}{0}
\renewcommand{\theequation}{\thesection.24\alph{equation}}
\ba
\lb{dEF}
&&d_i = q^{\Lambda_{i-1} - \Lambda_{i}}\quad (i = 1, \dots , n \; , \;\;
\Lambda_{0} = 0 = \Lambda_{n} ) \; ,\nonumber\\
&&e_i=({\bar q}-q) E_i \, ,
\quad f_i=({\bar q}-q)F_i \,\\
&& (\bq -q) f_{12} = f_2 f_1 - q f_1 f_2 =
(\bq -q)^2 (F_2 F_1 - q F_1 F_2) \  { etc.,}\nonumber\\
&& (\bq -q) e_{21} = e_1 e_2 - q e_2 e_1 =
(\bq -q)^2 (E_1 E_2 - q E_2 E_1) \ { etc.\;.}
\ea
Here $\Lambda_{i}$ are the fundamental co-weights of $sl(n)$
(related to the co-roots $H_{i}$ by $H_{i} = 2\Lambda_{i} -
\Lambda_{i-1} - \Lambda_{i+1}$); $E_i\,$ and
$F_{i}\,$
are the raising and lowering operators satisfying
\setcounter{equation}{0}
\renewcommand{\theequation}{\thesection.25\alph{equation}}
$$
[ E_{i}, \, F_{j} ] = [H_{i}] \, \delta_{ij} \; \;
\left( [H]:= \frac{q^H-{\bar q}^H}{q-\bar q} \right)\,,
$$
\be
[E_i ,\, E_j ] = 0 = [F_i,\, F_j]\ \ \ {for}\ \ \ |j-i|\ge 2\,,
\ee
$$
q^{{\Lambda}_i}E_j=E_j\; q^{{\Lambda}_i +{\delta}_{ij}}\,,\quad \
q^{{\Lambda}_i}F_j=F_j\; q^{{\Lambda}_i -{\delta}_{ij}}\,,\nonumber
$$
\be
\left[ 2 \right] X_i X_{i\pm1}X_i =X_{i\pm 1}X_i^2 +X_i^2X_{i\pm1}\ \
\ for\ \ \ X=E,\;F\,.
\ee
We note that the invariance under the coaction of $SL_q(n)$ (1.22a)
is, in effect, equivalent to the covariance relations
\setcounter{equation}{25}
\renewcommand{\theequation}{\thesection.\arabic{equation}}
\be
\lb{1.26}
\begin{array}{c}
q^{H_{i}} \, u_{\a}(x) \, {\bar q}^{ H_{i}} =
q^{\delta^{i}_{\a} - \delta^{i+1}_{\a}} \, u_{\a}(x) \; , \;\;
[E_i ,\, u_{\a}]={\delta}^{i+1}_{\a}u_{\a -1}(x)q^{H_i}\;, \;\;
\\ \\
F_iu_\a (x)-q^{{\delta}_{\a}^{i+1}-{\delta}_{\a}^i}u_{\a}(x)F_i
={\delta}_{\a}^iu_{\a +1}(x)\;.
\end{array}
\ee

\vspace{5mm}

\subsection{$R$-matrix realizations of the Hecke algebra;
quantum antisymmetrizers}

\medskip

The $R$-matrix for the quantum deformation of any (simple) Lie
algebra can be obtained as a representation of Drinfeld's
universal $R$-matrix \cite{D}. In the case of the defining
representation of $SU(n)$ the braid
operator (1.9) gives rise, in addition, to a representation of
the {\em Hecke algebra}. This fact, exploited in \cite{HIOPT}, is
important for our understanding of the dynamical $R$-matrix. We
recall the basic definitions.

For any integer $k\ge 2$ let $H_k(q)$ be an associative
algebra with generators $1, g_1,\ldots , g_{k-1}\,$, depending on
a non-zero complex parameter $q$, with defining relations
\setcounter{equation}{0}
\renewcommand{\theequation}{\thesection.27\alph{equation}}
\ba
&g_i g_{i+1} g_i = g_{i+1} g_i g_{i+1} &{\rm for}\  1\le i\le
k-2\ \ ({\rm if}\ k\ge 3)\,,\\
&g_i g_j = g_j g_i &{\rm for}\  |i-j|\ne 1, \ 1\le i, j \le k-1\,,\\
& g_i^2 = 1 + (q-\bq ) g_i &{\rm for}\  1\le i\le k-1,\ \  \bq := q^{-1}\,.
\ea
\setcounter{equation}{0}
\renewcommand{\theequation}{\thesection.28\alph{equation}}
The $SL(n)\,$ braid operator $\R\,$ (see (1.8b,c), (1.9a))
generates a representation
$\rho_n: H_k(q)\rightarrow End(V^{\otimes k})$, $V={\C}^n$ for any
$k\geq 2\,,$
\be
\rho_n(g_i)= q^{1\over n} {\R}_{i i+1}\quad {\rm or} \quad
[\rho_n(g_i)]^{\pm 1} = q^{\pm 1}\1 - A_i\,,
\ee
where $A$ is the $q$-antisymmetrizer
\be
\ \ A^{\a_1 \a_2}_{\b_1 \b_2} =
q^{\epsilon_{\a_2 \a_1}} \delta^{\a_1 \a_2}_{\b_1 \b_2} -
\delta^{\a_1 \a_2}_{\b_2 \b_1}\,, \ \
q^{\epsilon_{\a_2\a_1}} =
\left\{
\begin{array}{ccc}
\bq & {\rm for} & \a_1 > \a_2 \\
1 & {\rm for} & \a_1 = \a_2 \\
q & {\rm for} & \a_1 < \a_2
\end{array}
\right .
\ee
($A_i=[2]A^{(i+1,i)}$ in the -- suitably extended --
notation of \cite{HIOPT}; note that
for $q^2=-1\,, \, [2]=0\,$ the "normalized antisymmetrizer"
$A^{(i+1,i)}$ is ill defined while $A_i$ still makes sense).
Eqs.(1.27) are equivalent to the following
relations for the antisymmetrizers $A_i\,$:
\setcounter{equation}{0}
\renewcommand{\theequation}{\thesection.29\alph{equation}}
\ba
&&A_i A_{i+1} A_i - A_i = A_{i+1} A_i A_{i+1} -
A_{i+1}\\
&&A_i A_j = A_j A_i \quad {\rm for} \ \ |i-j|\ne 1\\
&&A_i^2 = [2] A_i\,.
\ea

\vspace{5mm}
\noindent
{\em Remark 1.3~} We can define (see, e.g. \cite{Gur}) the
higher antisymmetrizers $A_{i\,j}\,,\ i<j\,$ inductively, setting
\setcounter{equation}{0}
\renewcommand{\theequation}{\thesection.30\alph{equation}}
\ba
\lb{anti0}
&&A_{i\,j+1}:=A_{i\,j}(q^{j-i+1} - q^{j-i}\rho_n(g_j)+\dots +
(-1)^{j-i+1}\rho_n(g_j g_{j-1}\dots g_i)) =
\nonumber\\
&&= A_{i+1\,j+1}(q^{j-i+1} - q^{j-i}\rho_n(g_i)+\dots +
             (-1)^{j-i+1}\rho_n(g_i g_{i+1}\dots g_j)).
\ea
They can be
also expressed in terms of antisymmetrizers only:
\ba
\lb{anti}
&&A_{i\, i+1}=
A_i\,,\ \
A_{i\, j+1} =
{1\over{[j-i]!}}\left( A_{i\,j} A_j A_{i\,j} - [j-i][j-i]!
A_{i\,j} \right) =\nonumber\\
&&= {1\over{[j-i]!}}\left(
A_{i+1\, j+1} A_i A_{i+1\, j+1} - [j-i][j-i]! A_{i+1\,
j+1} \right) \,.
\ea

\vspace{5mm}

The term "$q$-antisymmetrizer" is justified by the relation
\setcounter{equation}{0}
\renewcommand{\theequation}{\thesection.31\alph{equation}}
\be
\left( \rho_n(g_i)+\bq\right) A_{1\,j}\, =\, 0\, =\,
A_{1\,j} \left( \rho_n(g_i)+\bq\right)
\ee
or
\be
A_{1\,i} A_{1\,j} = [i]!\, A_{1\,j} \quad {\rm for}\quad 1<i\le
j\,.
\ee
The dependence of the representation $\rho_n\,$ on $n$ (for
$G=SU(n)\,$) is manifest in the relations
\setcounter{equation}{31}
\renewcommand{\theequation}{\thesection.\arabic{equation}}
\be
\lb{rank}
A_{1\,n+1} = 0\,, \quad {\rm rank}\, A_{1\,n} = 1\,.
\ee
$A_{1\,n}$ can be written as a
(tensor) product of two Levi-Civita tensors:
\be
\lb{2e}
A_{1\,n} = \Eup\, \Edo\,, \quad \Edo\, \Eup = [n]!\,,
\ee
the second equation implying summation in all $n$ repeated indices.
We can (and shall) choose the covariant and the contravariant ${\cal
E}$-tensors equal,
\be
\lb{e}
{\cal E}^{\scriptscriptstyle \a_1 \a_2 \dots \a_n} =
{\cal E}_{\scriptscriptstyle \a_1 \a_2 \dots \a_n} =
\bq^{n(n-1)/4}\, (-q)^{\ell(\sigma)}\quad {\rm for} \quad
\sigma = {n_{~}, \dots ,1_{~}\choose \a_1, \dots ,\a_n}\,
\left( \,\in {\cal S}_n\, \right)\,;
\ee
here $\ell(\sigma)$ is the length of the permutation
$\sigma\,$. (Note the difference between (\ref{e}) and the
expression (2.5) of \cite{HIOPT} for ${\cal E}$ which can be traced
back to our present choice (1.28) for ${\rho}_n(g_i)\,$ --
our $\R_{i\, i+1}$ corresponding to $\R_{i+1\, i}$ of \cite{HIOPT} --
cf. Remark 1.1.)

We recall for further reference that the first equation
(\ref{rank}) is equivalent to either of the following two relations:
\setcounter{equation}{0}
\renewcommand{\theequation}{\thesection.35\alph{equation}}
\ba
\lb{AAA}
A_{1\,n} A_{2\, n+1} A_{1\, n} &=& ([n-1]!)^2 \, A_{1\, n}\\
A_{2\, n+1} A_{1\, n} A_{2\, n+1} &=& ([n-1]!)^2\, A_{2\, n+1}
\ea
(see Lemma 1.1 of \cite{HIOPT}); this agrees with
(\ref{2e}), (\ref{e}) since
\setcounter{equation}{0}
\renewcommand{\theequation}{\thesection.36\alph{equation}}
\ba
\lb{EdEu}
\D2\,\Eup &=& (-1)^{n-1}\, [n-1]!\  \d^{|1{\cal i}}_{{\cal h}n+1|} \\
\Edo\,\U2 &=& (-1)^{n-1}\, [n-1]!\  \d^{|n+1{\cal i}}_{{\cal h}1|}\ .
\ea
\setcounter{equation}{36}
\renewcommand{\theequation}{\thesection.\arabic{equation}}
We shall encounter in Section 2 below another, "dynamical", Hecke
algebraic representation of the braid group which has the same
form (1.28a) but with a "dynamical antisymmetrizer", i.e., $A_i =
A_i (p)\,$, a (rational) function of the $q$-weights
$\left( q^{p_1},\ldots , q^{p_n} \right)$ which satisfies a
finite difference ("dynamical") version of (1.29a).

\vspace{5mm}

\subsection{Barycentric basis, shifted $su(n)$ weights; conformal
dimensions}

\medskip

Let $\{ v^{(i)}, \, i=1,\ldots , n\}$ be a symmetric {\em
"barycentric basis"} of (linearly dependent) real traceless
diagonal matrices (thus $\{ v^{(j)}\}\,$ span a real Cartan subalgebra
${\frak h} \subset  sl(n)$):
\be
\lb{bary}
(v^{(i)})^j_k = \left(\d_{ij} - {1\over n}\right)\d^j_k\quad\Rightarrow\quad
\sum_{i=1}^n\, v^{(i)} = 0\,,\quad
( v^{(i)} | v^{(j)} ) = \d_{ij} - \frac{1}{n}\,.
\ee
(The inner product of two matrices is given by the trace of
their product.) Analogously, the $n$ "barycentric"
components $p_i\,$ of a vector in the $n-1\,$ dimensional dual space
${\frak h}^*\,$ are determined up to a common additive constant and
can be fixed by requiring $\sum_{i=1}^n p_i = 0\,.$
Specifying thus the bases, we can make correspond to any such vector
in ${\frak h}^*\,$ a unique diagonal matrix $p \in \frak{h}\,,$
\be
\lb{p}
p=\sum_{i=1}^n p_i\, v^{(i)}\ =
\left(\matrix{ p_1 &0 &\ldots &0\cr
0 &p_2 &\ldots &0\cr
\ldots &\ldots &\ldots &\ldots\cr
0 &0 &\ldots &p_n\cr}\right) \,,\qquad \sum_{i=1}^n\, p_i = 0\,,
\ee
and vice versa. In particular, the simple $sl(n)$ roots $\a_i$ and the
fundamental $sl(n)$ weights $\Lambda^{(j)}\,,\ i,j=1,\dots ,n-1\,,$
satisfying $(\Lambda^{(j)} | \a_i ) = \d^j_i\,,$
correspond to the following diagonal matrices (denoted by the same symbols),
\be
\lb{roots}
\a_i=v^{(i)}-v^{(i+1)}\,,\quad\Lambda^{(j)} = \sum_{\ell =1}^j v^{(\ell )}
\equiv (1-\frac{j}{n}) \sum_{\ell =1}^{j} v^{(\ell )} -
\frac{j}{n} \sum_{\ell = j+1}^n v^{(\ell )}\,,
\ee
respectively. Expanding $p\,$ (\ref{p}) in the basis of fundamental weights,
$$
p\, =\,\sum_{i=1}^n  p_i\, v^{(i)} \, =\,
\sum_{j=1}^{n-1} p_{j\, j+1} \, \Lambda^{(j)}\,,\quad\quad p_{ij} := p_i
- p_j\,,$$
one can characterize a {\em shifted dominant weight}
\be
\lb{weight}
p=\Lambda +\rho\,,\quad
\Lambda = \sum_{i=1}^{n-1}\,\lambda_i\,\Lambda^{(i)}\,,\ \
\lambda_i\in {\Z}_+\,,\quad
\rho = \sum_{i=1}^{n-1}\,\Lambda^{(i)}\
= {1\over 2}\sum_{\a >0}\,\a
\ee
($\rho\,$ is the $sl (n)\,$ Weyl vector) by the relations
\be
\lb{dominant}
p_{i\, i+1} = \lambda_i +1 \in {\Bbb N}\,,\qquad
i=1,2,\ldots ,n-1\,.
\ee
The non-negative integers $\lambda_i = p_{i\, i+1} - 1\,$ count
the number of columns of length $i$ in the Young tableau that
corresponds to the IR of highest weight $p$ of $SU(n)$ -- see,
e.g., \cite{Ful}. Conversely, $p_i\,$ satisfying (\ref{p}) can be
expressed in terms of the integer valued differences $p_{ij}\,$ as
$p_i= {1\over n}\sum_{j=1}^n p_{ij}\,.$

Dominant weights $p$ also
label highest weight representations of $U_q\,\,.$ For integer
heights $h\, (\ge n )\,$ and $q$ satisfying (\ref{0.1}) these are
(unitary) irreducible if $( n-1\le~)\ p_{1 n}\, \le h\,$. The
{\em quantum dimension} of such an IR is given by (see, e.g.,
\cite{CP})
\be
\lb{qdim}
d_q(p) = \prod_{i=1}^{n-1} \{\, {1\over{[i]!}}\,\prod_{j=i+1}^n\,
[p_{i j}]\}\quad (\,\ge 0\quad{\rm for}\quad
p_{1 n} = p_1 - p_n \,\le h )\,.
\ee
For $q\rightarrow 1\ (h\rightarrow \infty )\,, \ [m]\rightarrow m\,$
we recover the usual (integral) dimension of the IR under consideration.

The {\em chiral observable algebra} of the $SU(n)$ WZNW model is
generated by a local current $j(x)\in su(n)\,$ of height $h\,$. In
contrast to gauge dependent charged fields like $u(x)\,$, it is
periodic, $\, j(x+2\pi ) = j(x)\,$. The quantum version of the
classical field-current relation $ i\, j(x) = k\, u'(x)
u^{-1}(x)\,$ is the operator Knizhnik-Zamolodchikov equation
\cite{KZ} in which the level $k$ gets a quantum correction (equal
to the dual Coxeter number $n$ of $su(n)\,$):
\be
\lb{KZ}
h\, u'(x) = i\, :j(x)\, u(x) :\,,\quad h=k+n\,.
\ee
Here the normal product is defined in terms of the current's
frequency parts
\be
\lb{normal}
:j\, u: = j_{(+)}\,u + u\, j_{(-)}\,,\quad
j_{(+)} (x) = \sum_{\nu=1}^\infty\,J_{-\nu} e^{i\nu x}\,,\quad
j_{(-)} (x) = \sum_{\nu =0}^\infty\,J_\nu e^{-i\nu x}\,.
\ee
The canonical chiral stress energy tensor and the conformal
energy $L_0\,$ are expressed in terms of $j$ and its modes by the
Sugawara formula:
\be
\lb{sugawara}
{\cal T} (x)={1\over {2 h}}\,{\rm tr}\, :j^2 : (x)\ \Rightarrow \,
L_0 = \int_{-\pi}^{\pi} {\cal T} (x) {{dx}\over{2\pi}} =
{1\over {2h}}\,{\rm tr}
\left( J_0^2 + 2\sum_{\nu =1}^{\infty} J_{-\nu} J_\nu\,\right)\,.
\ee
Energy positivity implies that the state space of the chiral
quantum WZNW theory is a direct sum of (height $h\,$) ground state modules
${\cal H}_p\,$ of the Kac-Moody algebra
${\widehat{su}}(n)$ each entering with a finite multiplicity:
\be
\lb{space}
{\cal H}\, =\, \bigoplus_p\,{\cal H}_p\otimes {\cal F}_p\,,\quad
{\rm dim}\,{\cal F}_p < \infty\,.
\ee

We are not fixing at this point the structure of the internal
spaces ${\cal F}_p\,.$ In the simpler but unrealistic case of
generic $q\,$ explored in Section 3.1 each ${\cal F}_p\,$ is an
irreducible $U_q\,$ module and the direct sum $\oplus_p\;{\cal
F}_p\,$ carries a Fock type representation of the intertwining
quantum matrix algebra ${\cal A}\,$ introduced below. The
irreducibility property fails, in general, for $q\,$ a root of
unity (as discussed in Section 3.3). It is conceivable that in
this (realistic) case the label $p\,$ should be substituted by
the set of eigenvalues of the $U_q\,$ Casimir operators which are
symmetric polynomials in $q^{p_i}\,.$ (In the case of
$U_q(sl_2)\,$ the single Casimir invariant depends on $q^p +
\bq^p\,,\ p\equiv p_{12}\,,$ which suggests that $p\,$ and
$2h-p\,$ should refer to the same internal space.)

Each ${\cal H}_p\,$ in the direct sum (\ref{space}) is a positive
energy graded vector space, \be \lb{grad} {\cal H}_p \, =\,
\oplus_{\nu =0}^\infty\; {\cal H}_p^\nu\,,\quad \left( L_0 -
\Delta (p) - \nu\right)\; {\cal H}_p^\nu\, =\, 0\,, \quad {\rm
dim} {\cal H}_p^\nu < \infty\,. \ee It follows from here and from
the current algebra and Virasoro CR \be \lb{CR} [ J_\nu , L_0 ] =
\nu J_\nu \,,\quad [L_\nu , L_0 ] = \nu L_\nu \quad (\nu \in \Z )
\ee that $ J_\nu {\cal H}_p^0 = 0 = L_{\nu} {\cal H}_p^0\quad {\rm
for}\quad \nu = 1,2,\dots .\,$ Furthermore, ${\cal H}_p^0\,$ spans
an IR of $su(n)\,$ of (shifted) highest weight $p\,$ and dimension
$d_1(p)\,$ (the $q \to 1\,$ limit of the quantum dimension
(\ref{qdim})). The {\em conformal dimension} (or {\em conformal
weight}) $\Delta (p)\,$ is proportional to the ($su(n)$-) second
order Casimir operator $|p|^2 - |\rho |^2\,:$
\be
\lb{Delta}
2h\Delta (p) = |p|^2 - |\rho |^2 = {1\over n} \sum_{1\le i < j\le
n}\,p_{i j}^2 - {{n(n^2-1)}\over {12}}\,.
\ee
Note that the conformal dimension
$\Delta{(p^{(0)})}\,$ of the trivial representation \be \lb{triv}
p^{(0)} \; =\; \{ p\;;\; p_{i i+1} = 1\;,\ 1\le i\le n-1 \;\} \ee
is zero. This follows from the identity
\ba
\lb{idtriv}
&&n |p^{(0)}|^2 = \sum_{i=1}^{n-1}\; \sum_{j=i+1}^n\; (p_{ij}^{(0)})^2 =\nonumber\\
&&= \sum_{i=1}^{n-1}\; \sum_{j=i+1}^n\; (j-i)^2 =
\sum_{i=1}^{n-1}\; {{n-i}\over 6} (2n-2i+1)(n-i+1) =\nonumber\\
&&= {{n^2(n^2-1)}\over{12}} = n |\rho |^2
\quad \Rightarrow \quad |p^{(0)}|^2 - |\rho |^2 = 0\,.
\ea
The eigenvalues of the braid operator $\R$ (1.9), (1.28)
are expressed as products of exponents of conformal
dimensions. Let indeed $p^{(1)}\,$ be the weight of the defining
$n$-dimensional IR of $su(n)\,$:
\be
\lb{quark}
p_{12}^{(1)}\, =\, 2\,,\quad p_{i i+1}^{(1)} \, =\, 1\ \
{\rm for}\ \ i\ge 2
\ee
while $p^{(s)}$ and $p^{(a)}$ be the weights of the symmetric and the
antisymmetric squares of $p^{(1)}\,$, respectively,
\ba
&&p_{1 2}^{(a)}\, = \, 1\ (\, =\, p_{i i+1}^{(a)}\
{\rm for}\ i\ge 3\,)\,,\quad
p_{2 3}^{(a)}\, = \, 2\quad (\,{\rm for}\ n\ge 3\, )\nonumber\\
&&p_{1 2}^{(s)}\, = \, 3\,,\ \ \ p_{i i+1}^{(s)}\, =\, 1\ \ \,
{\rm for}\ n-1\ge i\ge 2\,.
\lb{as}
\ea
The corresponding conformal dimensions
$\Delta_{~}\, = \Delta (p^{(1)})\,,\
\Delta_a\, = \Delta (p^{(a)})\,$ and
$\Delta_s\, = \Delta (p^{(s)})\,$ are computed from
(\ref{Delta}):
\ba
\lb{confdim}
2h\,\Delta\, &=&\, |p^{(1)}|^2 - |\rho |^2
\, =\, {{n^2-1}\over n}\,,{\nonumber}\\
2h\,\Delta_a\, &=&\, |p^{(a)}|^2 - |\rho |^2
\, =\, 2\, {{n+1}\over n} (n-2)\,,\\
2h\,\Delta_s\, &=&\, |p^{(s)}|^2 - |\rho |^2
\, =\, 2\, {{n-1}\over n} (n+2)\,{\nonumber}.
\ea
The two eigenvalues of $\R\,$ (evaluated from the nonvanishing
$3$-point functions that involve two fields $u(x)\,$ (or
$\varphi (z)\,$ -- see (\ref{analytic})) of conformal weight
$\Delta\,,$
\be
\lb{eigenvalues}
e^{i\pi\,(2\Delta -\Delta_s )} = q^{{n-1}\over n}\,,\quad
-\, e^{i\pi\,(2\Delta -\Delta_a )} = -\, {\bq}^{{n+1}\over n}\,,
\ee
appear with multiplicities
\be
\lb{multi}
d_s = {n+1\choose 2}\,,\quad d_a = {n\choose 2}\qquad (d_a+d_s =
n^2)\,,
\ee
respectively. The deformation parameter
\be
\lb{q}
q^{1\over n} \, = \, e^{-i\,{{\pi}\over nh}}\,,
\ee
computed from here, satisfies (\ref{0.1}) as anticipated. For
$q\,\to\, 1\,$ the eigenvalues (\ref{eigenvalues}) of $\R\,$ go
into the corresponding eigenvalues $\pm 1\,$ of the permutation
matrix $P\,$; furthermore,
\be
\lb{evP}
{\rm det}\,\R\, =\, {\rm det}\, P\, =\, (-1)^{d_a}\, = \,
\left\{
\begin{array}{ccc}
-1 & {\rm for} & n=2,3 \,{\rm mod}\, 4 \\
1 & {\rm for} & n = 0,1\, {\rm mod}\, 4 \\
\end{array}
\right .
\,.
\ee
Eq.(\ref{eigenvalues}) illustrates the early observation (see, e.g.,
\cite{A-GGS}) that the quantum group is determined by basic
characteristics (critical exponents) of the underlying conformal
field theory.

\vspace{5mm}

\section{CVO and $U_q$-vertex operators: monodromy and braiding
}
\subsection{Monodromy eigenvalues and ${\cal F}_p\,$ intertwiners
}

\medskip

\setcounter{equation}{0}
\renewcommand{\theequation}{\thesection.\arabic{equation}}

The labels $p$ of the two factors in each term of the expansion
(\ref{space}) have different nature. While ${\cal H}_p\,$ is a
ground state current algebra module for which $p\,$ stands for the
shifted weight
$p^{KM}\,$ (such that $(p_i^{KM}-p_i){\cal H}_p=0\;$)
of the ground state
representation of the ${\widehat{su}}(n)\,$
current algebra of minimal conformal
dimension (or energy) $\Delta (p),\ {\cal F}_p\,$ is a $U_q\,$
module (the quantum group commuting with the currents).
We introduce, accordingly, the field of
rational functions of the commuting operators
$q^{\p_i}\,$ (giving rise to an abelian group and) such that
\be
\lb{q^p}
\prod_{i=1}^n\, q^{\p_i}\, = \, 1\,,\qquad\
\left(\,q^{\p_i}\, -\,q^{p_i}\right)\, {\cal F}_p\,=\,0\,,\qquad\
[q^{\p_i}\,,\, j(x)]\, =\,0
\ee
with $p_{i j}\,$ obeying the condition
(\ref{dominant}) for dominant weights.

We shall split the $SU(n)\times U_q(sl_n)\,$ covariant field $u(x) =
\left( u^A_\a (x)\right)\,$ into factors which intertwine
separately different ${\cal H}_p\,$ and ${\cal F}_p\,$ spaces.

A CVO $u_j(x,p)\,$ (for $p\equiv p^{KM}\;$)
is defined as an intertwining map between ${\cal H}_p\,$ and
${\cal H}_{p+v^{(j)}}\,$ (for each $p$ in the sum (\ref{space})).
Noting that ${\cal H}_p\,$ is an eigenspace of $e^{2\pi i\, L_0}\,$,
\be
\lb{specL}
{\rm Spec}\, L_0\, |_{{\cal H}_p}\,\subset \Delta_h(p)\, +{\Z}_+ \quad
\Rightarrow\quad \{ {e}^{2\pi i\, L_0}\, - \,
{e}^{2\pi i\,\Delta_h(p)}\,\}\, {\cal H}_p\, = \, 0\,,
\ee
we deduce that $u_j(x,p )\,$ is an eigenvector of the monodromy
automorphism:
\setcounter{equation}{0}
\renewcommand{\theequation}{\thesection.3\alph{equation}}
\be
\lb{monodromyauto}
u_j(x+2\pi ,p)=e^{2\pi iL_0}\, u_j(x,p)\,e^{-2\pi iL_0}\,=
u_j(x,p)\,\mu_j(p)
\ee
where, using (\ref{Delta}) and the relation
$(p\, |\, v^{(j)} )\,=\,p_j\,,$ we find
\be
\lb{mu}
\mu_j(p) \,:=\, e^{2\pi i \{
\Delta_h(p+v^{(j)}) - \Delta_h(p)
\}}\,=\,
q^{{1\over n}-2p_j -1}\,.
\ee
The monodromy matrix (\ref{1.5}) is diagonalizable whenever its
eigenvalues (2.3b) are all different. In particular, for the "physical
IR", characterized by $p_{1n}\, <\,h\,,\ M\,$ is diagonalizable.
The exceptional points are those $p$ for which there exists a pair
of indices $1\le i <j\le n\,$ such that $q^{2\,p_{i j}}\, =\, 1\,$,
since we have
\be
\lb{exceptional}
{{\mu_i(p)}\over{\mu_j(p)}}\,=\,q^{-2\,p_{i j}}
\,.
\ee
According to (\ref{qdim}) all such "exceptional" ${\cal F}_p\,$
have zero quantum dimension ($[p_{ij}]=0$).

\vspace{5mm}
\noindent
{\em Remark 2.1~} The simplest example of
a non-diagonalizable $M\,$ appears for $n=2\,, \ p \ (\equiv
p_{12} ) = h\,$ when $\mu_1(h) = - \bq^{1\over 2} = \mu_2(h)\,.$
In fact any ${\widehat{su}}(2)\,$ module ${\cal H}_{h-\ell}
\,, \ 0\le\ell\le h-1\,$ contains a singular (invariant) subspace
isomorphic to ${\cal H}_{h+\ell}\,$
\cite{Kac}; note that, for $p=h-\ell\,,\ \tilde p = h+\ell\,,$
\be
\mu_1(p) = -q^{\ell-{1\over 2}} = \mu_2(\tilde p)\,,\quad
\mu_2(p) = -\bq^{\ell+{1\over 2}} = \mu_1(\tilde p)
\ee
(cf. (2.3b)).
It turns out that, in general,
\be
\mu_1(p) = \mu_{n}(\tilde p)\,,\quad
\mu_n(p) = \mu_{1}(\tilde p)\,,\quad
\mu_i(p) = \mu_{i}(\tilde p)\,,\ \, i=2,3,\dots ,n-1\,.
\ee
Indeed, if
$|{\rm hwv}{\cal i}_p\,$ is the highest weight vector
in the minimal energy subspace
${\cal H}_p^0\,$ of the $\widehat{su}(n)\,$ module ${\cal H}_p\,$
and $\theta = \a_1+\dots\a_{n-1}\,$ is the $su(n)\,$ highest root, then the
corresponding singular vector can be written in the form \cite{GW}
\setcounter{equation}{3}
\renewcommand{\theequation}{\thesection.\arabic{equation}}
\ba
\lb{singvect}
&&\left( E_{-1}^{\theta} \right)^{h-p_{1 n}} |{\rm hwv}
{\cal i}_p\  \sim\ |{\rm hwv}
{\cal i}_{\tilde p}\,,\nonumber\\
&&{\tilde p}_1=h+p_n\,,\quad {\tilde p}_n=-h+p_1\,,\quad
{\tilde p}_i = p_i\,,\ \, i=2,3,\dots ,n-1\,,\\
&&\Delta ({\tilde p}) - \Delta (p) = h-p_{1n} \in \Z\,.\nonumber
\ea
To prove
(2.3e) one just has to insert ${\tilde p}_i\,$ into (2.3b).
One concludes that the monodromy $M\,$ always has coinciding
eigenvalues on ${\cal F}_p \oplus {\cal F}_{\tilde p}\,$
(suggesting the inclusion of this direct sum into  a single
indecomposable $U_q\,$ module -- cf. \cite{FK}). The nondiagonalizability
of the monodromy matrix in the extended state space
may require a modification of the splitting (\ref{vIRF}) below for
$p_{1n}\ge h\,.$ The study of
this question is, however, beyond the scope of the present paper.

\vspace{5mm}

The CVO $u_j\,$ only acts on the factor ${\cal H}_p\,$ of the
tensor product ${\cal H}_p\otimes{\cal F}_p\,$ in (\ref{space});
hence, it shifts the Kac-Moody
operators $p_i^{KM}\,$ but not the
quantum group ones:
\be
\lb{new}
[ p_i^{KM}\,,\; u_j ] = (\d_{ij}-\frac{1}{n})\; u_j  \,,\quad
[q^{{\hat p}_i}\,,\; u_j ]\,=\,0\quad
\left( ( q^{p_i^{KM}} - q^{{\hat p}_i} )\;{\cal H} = 0\;\right)\,.
\ee
We shall skip from now on (as we did in
Eqs.(2.3),(\ref{singvect}))
the superscript {\footnotesize{$KM$}} of
the argument $p\,$ of the CVO $u_j\,$ as well as the hat on the
quantum group operators $q^{p_i}\,$ since the distinction
between the two labels $p$ should be
obvious from the context (and does
not matter when acting on the diagonal chiral state space
(\ref{space})).

The {\em intertwining quantum matrix algebra} ${\cal A}\,$
is generated by $q^{p_i}\,$ and by the elements of the matrix
$a\,=\, \left(a^j_\a\right)\,,\ j,\a=1,\ldots ,n\,$
which shift $p\,$ (commuting with $p^{KM}\,$),
\be
\lb{qa}
a^j_\a\,:\ {\cal F}_p\,\rightarrow\, {\cal F}_{p+v^{(j)}}\,,\quad
q^{p_i}\,a^j_\a\,=\,a^j_\a\,q^{p_i+\d_i^j-{1\over n}}\,.
\ee
The $SU(n)\times U_q(sl_n)\,$ covariant field
$u^A_\a (x)\,$ is related to the CVO $\; u^A_j(x,p)\,$ by the so called
vertex-IRF (interaction-round-a-face) transformation \cite{BBB}
\be
\lb{vIRF}
u^A_\a(x)\,=\,u^A_i(x,p)\otimes a^i_\a\,.
\ee
According to (\ref{vIRF}), $a^j_\a\,$ act on the second factor of
(\ref{space}) only and hence commute with the currents and
with the Virasoro generators.

It will be proven in Section 3.1 that the Fock space representation
of ${\cal A}\,$ in $\oplus_p {\cal F}_p\,$ for {\em generic} $q\,$
provides a model of $U_q\,.$
The exchange relations of $a^i_\a\,$ (displayed in Section 2.3 below),
combined with (\ref{vIRF}) and with the defining property
$a^j_\a\; |0{\cal i} = 0\,$ for $j>1\,$ of the vacuum vector
$|0{\cal i}\,$ (the unique normalized state in ${\cal F}_{p^{(0)}}\,$
for $p^{(0)}\,$ given by (\ref{triv})) yield, in particular, the
relation
\be
\lb{aV-a}
a^j {\cal F}_p = 0 \ \ {\rm for}\ \ j>1\ \ {\rm and}\ \  p_{j-1} = p_j+1\,.
\ee

The meaning of (\ref{qa}), (\ref{aV-a})
can be visualized as follows. To each finite dimensional representation
of $U_q\,$ with (dominant) highest weight $p\,$ we associate, as
usual, a {\em Young tableau}
$Y_{[\lambda_1 ,\dots ,\lambda_{n-1}]}\,$ with $\l_i\, ( = p_{i\;i+1} -1)\,$
columns of height $i\, (=1,2,\dots , n-1)\,.$ Then $a^j\,$ adds a
box to the $j^{\rm th}$ row of the
Young tableau of $p$ (provided $p_{j-1}>p_j +1$ for
$j=2,\dots ,n\,).$ Here are some examples for $n=4\;:$
$$
\\ \\ \\
$$
\begin{picture}(100,100)(-18,-50)
\put(10,25){$a^j\  |0{\cal i}\, =\, \delta^j_1$}
\put(76,25){\framebox(8,8)}
\put(91,25){;}
\put(138,25){$a^1$}
\put(153,34){\framebox(8,8)}
\put(161,34){\framebox(8,8)}
\put(153,26){\framebox(8,8)}
\put(153,18){\framebox(8,8)}
\put(178,25){$=$}
\put(196,34){\framebox(8,8)}
\put(204,34){\framebox(8,8)}
\put(212,34){\framebox(8,8)}
\put(196,26){\framebox(8,8)}
\put(196,18){\framebox(8,8)}
\put(225,25){,}
\put(290,25){$a^2$}
\put(305,34){\framebox(8,8)}
\put(313,34){\framebox(8,8)}
\put(305,26){\framebox(8,8)}
\put(305,18){\framebox(8,8)}
\put(326,25){$=$}
\put(344,34){\framebox(8,8)}
\put(352,34){\framebox(8,8)}
\put(352,26){\framebox(8,8)}
\put(344,26){\framebox(8,8)}
\put(344,18){\framebox(8,8)}
\put(366,25){,}
\put(50,-30){$a^3$}
\put(65,-21){\framebox(8,8)}
\put(73,-21){\framebox(8,8)}
\put(65,-29){\framebox(8,8)}
\put(65,-37){\framebox(8,8)}
\put(86,-30){$=\, 0\ ,$}
\put(189,-30){$a^4$}
\put(204,-21){\framebox(8,8)}
\put(212,-21){\framebox(8,8)}
\put(204,-29){\framebox(8,8)}
\put(204,-37){\framebox(8,8)}
\put(225,-30){$=\, c\,\,{\cal E}_{{\cal h}1234|}$}
\put(285,-30){\framebox(8,8)}
\put(304,-30){.}
\end{picture}
$$
\\ \\
$$

The exchange relations of $a^i_\a\,$ with the Gauss components
(\ref{1.5}) of the monodromy are dictated by
(\ref{MPu}),
\be
\lb{MPa}
M_{1\pm}\,P\,a_1\,=\,a_2\,\R^{\mp 1}\,M_{2\pm}
\ee
and reflect, in view of (\ref{dEF}), the $U_q\,$ covariance of
$a\,$:
\setcounter{equation}{0}
\renewcommand{\theequation}{\thesection.10\alph{equation}}
\ba
&&[E_a,\,a^i_\a ]\,=\,\d_{a\,\a-1}\,a^i_{\a-1}\,q^{H_a}\,,\quad
\raisebox{2pt}{${}_a$}
=1,\ldots ,n-1 \,,\\
&&[q^{H_a} F_a ,\,a^i_\a ] \,=\,\d_{a\,\a}\, q^{H_a} a^i_{\a+1}\,,\\
&&q^{H_a}\,a^i_\a\,=\,a^i_\a\,q^{H_a+\d_{a\,\a}-\d_{a\,\a-1}}\,.
\ea The transformation law (2.10) expresses the coadjoint action
of $U_q\,$. Comparing (\ref{1.2}), (2.3) and (\ref{vIRF}) we
deduce that the zero mode matrix $a\,$ diagonalizes the monodromy
(whenever the quantum dimension (\ref{qdim}) does not vanish);
setting
\setcounter{equation}{0}
\renewcommand{\theequation}{\thesection.11\alph{equation}}
\be
a\,M\,=\,M_p\,a
\ee
we find (from the above analysis
of Eq.(2.3)) the implication
\be
d_q(p)\,\ne\,0\quad\Rightarrow\quad
\left(M_p\right)^i_j\,=\,\d^i_j\, \mu_j(p - v^{(j)})\,, \quad
\mu_j(p - v^{(j)}) = q^{-2{p}_j+1-{1\over n}}\,.
\ee
It follows
from (2.11) that the subalgebra of ${\cal A}\,$ generated by the
matrix elements of $M\,$ commutes with all $q^{p_i}\,.$ As
recalled in (\ref{1.5}), (\ref{1.6}) and (1.24), the Gauss
components of $M\,$ are expressed in terms of the $U_q\,$
generators. We can thus state that the centralizer of $q^{p_i}\,$
in ${\cal A}\,$ is compounded by $U_q\,$ and $q^{p_i}\,.$

\vspace{5mm}

\subsection{Exchange relations among zero modes from braiding properties
of $4$-point blocks}

\medskip
\setcounter{equation}{11}
\renewcommand{\theequation}{\thesection.\arabic{equation}}

The exchange relations (\ref{uuR2}) for $u\,$ given by (\ref{vIRF}) can be
translated into quadratic exchange relations for the "$U_q\,$
vertex operators" $a^i_\a\,$ provided that the CVO $u(x,p )\,$
satisfy standard braid relations: if $0<x-y< 2\pi\,,$ then
\be
\lb{uuRp}
\Pi_{xy}\; u^B_i(y,p +v^{(j)})\,
u^A_j(x,p)\, =\,
u^A_k(x,p +v^{(l)})\,
u^B_l(y,p ) \, \R (p)^{kl}_{ij}\ ;
\ee
if $0<y-x<2\pi\,,$ then $\R (p)\,$ in the right hand side should
be substituted by the inverse matrix (cf. (1.9b)).
Indeed, consistency of (\ref{uuRp}) with (1.7d)
on the diagonal state space ${\cal H}\,$ (\ref{space})
requires that
\be
\lb{Rpaa=aaR}
\R (p )^{\pm 1}\, a_1\, a_2\, =\, a_1\, a_2\, \R^{\pm 1}\,,
\ee
where $p\,$ in $\R (p)\,$ should be understood as an operator,
see (\ref{q^p}) and (\ref{qa}).

It has been proven in \cite{HST} that Eq. (\ref{uuRp}) is in fact a
consequence of the properties of the chiral 4-point function
\be
\begin{array}{c}
\lb{4p}
w^{(4)}_{p'\,p}\,=\,
{\cal h} 0 | \,{\phi}_{{p'}^*}(z_1)
\,{\varphi}(z_2)\,{\varphi}(z_3)\,{\phi}_p(z_4)\, | 0 {\cal i}
\,=\\ \\
= \ \sum\limits_{i,j}\,{\cal S}^{i j}(p)\,s_{i j}(z_1,\ldots ,z_4;\,p)
\d_{p',p+v^{(i)}+v^{(j)}}
\end{array}
\ee
(we assume that the vacuum vector is given by the tensor
product of the vacuum vectors for the affine and quantum matrix
algebras). Here $\phi_p(z)\,$ and ${\phi}_{{p'}^*}(z)\,$ are
general $z$-picture primary chiral fields of weights $p\,$ and
${{p'}^*}\,,$ respectively, where ${{p}^*}\,$ is the weight conjugate to $p\,,$
\be
\lb{p^*}
p\ \rightarrow\ p^*\,=\,\{\,p^*_i\,=\,-p_{n+1-i} \} \quad \
\Leftrightarrow\ p^*_{i\,i+1}\,=\,p_{n-i\,n+1-i}\,,
\ee
$\varphi (z)\,$ is the "step operator"
(\ref{analytic}) (of weight $p^{(1)}\,$, see (\ref{quark}), i.e.,
$\varphi (z) \equiv \phi_{p^{(1)}} (z)\,$),
${\cal S}^{i j}(p)\,$ is the zero mode correlator
\be
\lb{Sij}
{\cal S}^{i j}(p)\,:=\,{\cal h} p+v^{(i)}+v^{(j)}\,|\,a^i\,a^j\,|\,p\,
{\cal i}\,,
\ee
while $s_{i j}\,$ is the conformal block expressed in terms of
a function of the cross ratio $\eta\,$:
\be
\begin{array}{c}
\lb{sij}
s_{i j}(z_1,\,z_2,\,z_3,\,z_4;\,p)\, : =\,\\ \\
=\,{\cal h} 0 | \,{\phi}^{p^{(0)}}_{{p'}^*}(z_1, p')\,
{\varphi}_i(z_2,\,p+v^{(j)})\,
{\varphi}_j(z_3,\,p)\,{\phi}^p_p(z_4 , p^{(0)})\, | 0 {\cal i}\,=\\ \\
= D_{ij} (z_1, z_2, z_3, z_4 ;\, p )\,
f_{i j} (\eta ,\,p)\,.\\ \\
\end{array}
\ee
Here we use the standard notation
$$
\phi^{p_2}_p (z , p_1)\, :
{\cal H}_{p_1} \
\stackrel{\phi_p}{\longrightarrow}
\ {\cal H}_{p_2}
$$
for a CVO of weight $p\,$
(so that
$\varphi_\ell (z , p) \equiv \phi_{p^{(1)}}^{p+v^{(\ell )}} (z , p)\,$
is the $z$-picture counterpart of $u_\ell (x , p )\,$),
\ba
&&D_{ij} (z_1, z_2, z_3, z_4 ;\, p )\,
=\,{\left({ {z_{24}}\over{z_{12} z_{14}} }\right)}^{\Delta (p')}
{\left({{z_{13}}\over{z_{14} z_{34}}}\right)}^{\Delta (p)}\,
z_{23}^{-2\Delta}\,\eta^{\Delta_j-\Delta}\,
(1-\eta )^{\Delta_a}\,,\nonumber \\
&&\eta = {{z_{12} z_{34}}\over{z_{13} z_{24}}}\,,
\qquad\ \  \ p'\,=\,p+v^{(i)}+v^{(j)}\,;\nonumber
\ea
$\Delta (p)\,$ is given by (\ref{Delta}) and
$\Delta\,=\,\Delta (p^{(1)})\,=\, {{n^2-1}\over{2hn}}\,,
\ \Delta_j\,=\,\Delta (p+v^{(j)})\,,\ \Delta_a = \frac{(n+1)(n-2)}{2hn}\,$
($\Delta_a\,$ is the dimension (\ref{confdim})
of the antisymmetric tensor representation of weight $p^{(a)}\,$ in
(\ref{as})).
We are omitting here both $SU(n)\,$ and $SL_q(n)\,$ indices:
$s_{i j}\,$ (\ref{sij}) (and hence $f_{ij}\,$) is an $SU(n)\,$
invariant tensor in the tensor product of four IRs, while
${\cal S}^{ij}(p)\,$
(\ref{Sij}) is an $SL_q(n)\,$ invariant tensor. Only terms for
which both $p+v^{(j)}\,$ and $p+v^{(i)}+v^{(j)}\,$ are dominant weights
contribute to the sum (\ref{4p}). The prefactor
$D_{ij} (z_1 , z_2, z_3, z_4 ; p )\,$ in the right
hand side of (\ref{sij}) is fixed, up to a multiplicative
function of $\eta\,,$ by the M\"obius invariance of $f_{ij}\,.$
The choice of the powers of
$\eta$ and $1-\eta\,$ corresponds to extracting the leading singularities
(in both $s$- and $u$-channels) so that $f_{ij} (\eta ,\,p)\,$
should be finite (and nonzero) at $\eta = 0\,$ and $\eta = 1\,.$

We shall sketch the proof of (\ref{uuRp}); the
reader could find the details in \cite{HST} (see also \cite{TH}).

The conformal block $s_{ij}\,$ (\ref{sij}) is determined
as the $SU(n)\,$ invariant solution of the
Knizhnik-Zamolodchikov equation \cite{KZ}
\setcounter{equation}{0}
\renewcommand{\theequation}{\thesection.19\alph{equation}}
\be
\lb{kz-a}
\left( h{{\partial}\over{\partial z_2}}\,+\,
{ {{\cal C}_{12}}\over{z_{12}} }\,-\,
{ {{\cal C}_{23}}\over{z_{23}} }\,-\,
{ {{\cal C}_{24}}\over{z_{24}} }\,
\right)\, s_{i j} (z_1,\,z_2,\,z_3,\,z_4;\,p)\,=\,0
\ee
satisfying the above boundary conditions.
Inserting the expression (\ref{sij}) for $s_{ij}\,,$ one gets a
system of ordinary differential equations for the M\"obius
invariant amplitudes $f_{ij}\,$:
\be
\lb{kz-b}
\left( h {d\over{d \eta}}
- {{\Omega_{12}}\over{\eta}}
+ {{\Omega_{23}}\over{1-\eta}} \right)\, f_{ij} (\eta ;\, p)\, =\, 0\,.
\ee
Here $\,{\cal C}_{ab}
=\stackrel{\rightarrow}{t_a}.\stackrel{\rightarrow}{t_b}\,,
\ 1\le a<b\le 4\,$ is
the Casimir invariant of the corresponding tensor product of IRs of
$SU(n)\,;$ in our case $t_a\,,\ a=1,2,3,4\,$ generate the IRs of
weights ${p'}^*, p^{(1)}, p^{(1)}\,$ and $p\,,\,$ respectively.
The prefactor $D_{ij}\,$ being an $SU(n)\,$ scalar, $SU(n)\,$ invariance
of $s_{ij}\,$ implies
\be
\lb{kz-c}
\left( {\cal C}_{12} + {\cal C}_{23} + {\cal C}_{24} +
{{n^2-1}\over n} \right)\, f_{ij}\,=\,0\,,
\ee
so that
\be
\lb{kz-d}
\Omega_{12} = {\cal C}_{12} +
p_m + \d_{ij} +{{n^2+n-4}\over{2n}}\,,\quad
\Omega_{23} = {\cal C}_{23} +
{{n+1}\over n}\,,
\ee
where $m= {{\rm min}\, (i,j)}\,.$

Our objective is to study the braiding properties of the solution
$f_{ij}\,$ of (\ref{kz-b}) that is analytic in $\eta\,$ (and non-zero)
around $\eta = 0\,.$

It is important to observe that the space of invariant
$SU(n)\,$ tensors is in the case at hand at most two dimensional;
this allows us to find a convenient realization of the operators
$\Omega_{12}\,,\ \Omega_{23}\,$ \cite{CL, HST}. (In the $n=2\,$ case \cite{ZF, CF, STH}
this can be done even for four general isospins, due to the simple
rules for tensor multiplication in the $SU(2)\,$ representation ring.)

The existence of a solution of (2.19)
is guaranteed whenever the quantum dimension (\ref{qdim}) for each weight
encountered in (\ref{4p}) is positive,
\setcounter{equation}{19}
\renewcommand{\theequation}{\thesection.\arabic{equation}}
\be
\lb{sol}
n-1\,\le\, p_{1 n}\,,\ (p+v^{(j)})_{1n}\,,\ {p'}_{1 n}\,
<\,h\,,\qquad\
p' \equiv p+v^{(i)}+v^{(j)} \,.
\ee

In fact, for fixed $p\,$ and $p'\,$ in (\ref{sij})
and $i\ne j\,$ the $2\times 2\,$ matrix system Eq. (2.19b) gives
rise to a hypergeometric equation.
Assume, in addition, that $p+v^{(i)}\,$ is also a dominant weight. Then
both $s_{i j}\,$ and $s_{j i}\,$ will satisfy
Eq.(\ref{kz-a}) and provide a basis of independent solutions
of that equation (note that the sum in (\ref{4p}) reduces to two terms
with permuted $i\,$ and $j\,$).
More precisely, let $P_{23}\;{\Pi}_{23}\,s_{i j}
(z_1,\,z_3,\,z_2,\,z_4;\,p)\,$ be the analytic continuation of
$s_{i j}\,$ along a path $C_{23}\,$ obtained from $C_{12}\,$
(1.16a) by the substitution $1\to 2\,,\, 2\to 3\,$ (that is,
$C_{23}\,=\,\{ z_a(t) = e^{i\zeta_a(t)} \,,\ a=2,3\;;\,
\zeta_2(t)+\zeta_3(t)=x_2+x_3\,,\, \zeta_{23}(t)=
e^{-i\pi t} x_{23}\,,\ 0\le t \le 1\,\}$\,) with permuted $SU(n)\,$
indices $2$ and $3$. It satisfies again Eq.(\ref{kz-a}) and
hence is a linear combination of $s_{kl}
(z_1,\,z_2,\,z_3,\,z_4;\,p)\,$ with $(k,l)\,=\,(i,j)\,$ and
$(k,l)\,=\,(j,i)\,$:
\be
\lb{4pR}
P_{23}\;{\Pi}_{23}\, s_{i j} (z_1,\,z_3,\,z_2,\,z_4;\,p)\,=\,
s_{kl} (z_1,\,z_2,\,z_3,\,z_4;\,p)\,\DDR{kl}{ij}\,.
\ee
Here $\hat{R} (p)\;$ satisfies the {\it ice condition}: its
components $\DDR{kl}{ij}\,$ do not vanish only if the {\em
unordered} pairs $i,j\,$ and $k,l\,$ coincide -- i.e.,
\setcounter{equation}{0}
\renewcommand{\theequation}{\thesection.22\alph{equation}}
\be
\lb{DDR}
\DDR{kl}{ij}\,=\,
a^{kl}(p)\,\d^k_j\,\d^l_i\,+\,b^{kl}(p)\,\d^k_i\,\d^l_j\,.
\ee
Eq.(\ref{4pR}) is nothing but a matrix element version of
(\ref{uuRp}); hence, it yields the exchange relation
(\ref{Rpaa=aaR}) for $i\ne j\,$ ($\ \Rightarrow\ k\ne l\,$).

For $i=j\,$ the analytic continuation in the left hand side of
(\ref{4pR}) reduces to a multiplication by a phase factor.
In this case the space of $SU(n)\,$ invariant tensors is $1$-dimensional
(since the skewsymmetric invariant vanishes), and so is the space of
$U_q(s\ell_n )\,$ invariants.
The resulting equation for $f_{ii} (\eta \,;\, p) \,$ is first order:
$$
\left( h {d\over{d\eta}} + {2\over{1-\eta}}\right) f_{ii}(\eta\,;\, p )
= 0\,,
$$
and is solved by  $f_{ii}(\eta\, ;\, p) = c_{ii}(p)\, (1-\eta )^{2\over
h}\,.$
Substituting
$$ z_{23}\ \to\ e^{-i\pi} z_{23}\quad\Rightarrow\quad 1-\eta \ \to\
e^{-i\pi} \frac{1-\eta}{\eta}\,,\quad D_{ii}\ \to\ \bq^{\frac{n+1}{n}}
\eta^{\frac{2}{h}}\, D_{ii}\,,
$$
we get
$$s_{ii}\ \stackrel{\curvearrowright}\to\ q^{1-\frac{1}{n}}
s_{ii}\,.$$

Explicitly, the ($4\times 4\,$) $(i, j )$-block of ${\hat R} (p)\,$ has
the form
\be
\lb{Rp4}
\R^{(i,j)} (p_{ij} )\, =\,
\bq^{1\over n}
\left(\matrix{
q&0&0&0\cr 0&\frac{q^{p_{ij}}}{[p_{ij} ]}&{{[p_{ij} -1]}\over{[p_{ij} ]}}
\a (p_{ij} )&0\cr 0&{{[p_{ij} +1]}\over{[p_{ij} ]}}\a (- p_{ij} )&
-\frac{\bq^{p_{ij}}}{[p_{ij} ]}&0\cr 0&0&0&q}\right) \,, \ \ \a (p)\a
(-p) = 1\,,
\ee
i.e. (cf. (\ref{DDR}))
$$q^{\frac{1}{n}} a^{kl} (p_{kl}) = \a (p_{kl})
\frac{[p_{kl}-1]}{[p_{kl}]}\,,\quad
q^{\frac{1}{n}} b^{kl} (p_{kl}) = \frac{q^{p_{kl}}}{[p_{kl}]}\qquad{\rm
for}\quad k\ne l\,.$$
The arbitrariness reflected by $\a (p)\,$ is related to the
freedom of choosing the normalization of the two independent
solutions of the hypergeometric equation.

The matrix (\ref{Rp4}) coincides with the one, obtained independently in
\cite{HIOPT} by imposing consistency conditions on the intertwining
quantum matrix algebra of $SL(n)\,$ type. We shall display the ensuing
properties of ${\hat R} (p)\,$ in the following subsection.

\vspace{5mm}

\subsection{The intertwining quantum matrix algebra}

\medskip

\setcounter{equation}{22}
\renewcommand{\theequation}{\thesection.\arabic{equation}}

Among the various points of view on the $U_q(sl_2)\,$ intertwiners
(or "$U_q\,$ vertex operators") $a_{\a}^i\,$ the one which yields
an appropriate generalization to $U_q(sl_n)\,$ is the so called
"quantum $6j\,$ symbol" $\Rp$-matrix formulation of \cite{AF, FG,
BF, BBB}. The $n^2\times n^2\,$ matrix $\Rp\,$ satisfies the {\em
dynamical Yang-Baxter equation} (DYBE) first studied in \cite{GN}
whose general solution obeying the ice condition was found in
\cite{I2}.

The associativity of triple tensor products of quantum matrices
together with Eq.(\ref{RRR}) for $\R\,$ yields the DYBE for
$\Rp\,$:
\be
\lb{QDYBE}
\DR{12}\, \R_{23}(p-v_1)\, \DR{12}\, =\,
\R_{23}(p-v_1)\, \DR{12}\, \R_{23}(p-v_1)
\ee
where we use again the succinct notation of Faddeev et al. (cf.
Section 1):
\be
\lb{R23}
\left( \R_{23}(p-v_1)\right)^{i_1 i_2 i_3}_{j_1 j_2 j_3} \, =\,
\d^{i_1}_{j_1}\,\R (p-v^{(i_1)})^{i_2 i_3}_{j_2 j_3}\,.
\ee
In deriving (\ref{QDYBE}) from (\ref{Rpaa=aaR}) we use (\ref{qa}).
(The DYBE (\ref{QDYBE}) is only {\em sufficient} for the
consistency of the quadratic matrix algebra relations
(\ref{Rpaa=aaR}); it would be also necessary if the matrix $a\,$
were invertible -- i.e., if $d_q(p)\ne 0\,$.)

The property of the operators $\R_{i\, i+1}(p)\,$ to generate a representation
of the braid group is ensured by the additional requirement (reflecting
(1.27b))
\be
\lb{Rpvv}
\R_{12} (p+v_1+v_2 )\, =\, \R_{12}(p)\quad \Leftrightarrow\quad
\R^{ij}_{kl} (p) a^k_\a a^l_\b \, =\,
a^k_\a a^l_\b \,\R^{ij}_{kl} (p) \,.
\ee

The Hecke algebra condition (1.27c) for the rescaled matrices $\rho_n
(g_i)\,$ (1.28a) also fits our analysis of braiding properties of
conformal blocks displayed in the previous subsection.

It is not surprising that the direct inspection of the braiding
properties of the conformal blocks, from one side, and the common solution
of the DYBE,
(\ref{Rpvv}) and the Hecke algebra conditions \cite{I2, HIOPT}, from the
other, lead to the same result.
The solution (\ref{Rp4}) can be presented in a form similar to (1.28):
\be
\lb{A1}
q^{1\over n}\,\Rp\, =\, q\id - A(p)\,,\quad
A(p)^{ij}_{kl}\, =\, {{[p_{ij}-1]}\over{[p_{ij}]}}\, \left(
\d^i_k \d^j_l - \d^i_l\d^j_k\right)\,.
\ee
It is straightforward to verify the relations (1.29) for
$A_i(p) := q \id_{i\, i+1}- q^{1\over n} \DR{i\, i+1} \,$; in particular,
\be
\lb{A2}
[p_{ij}-1]\, +\, [p_{ij}+1]\, =\, [2]\,[p_{ij}]\quad\Rightarrow\quad
A^2 (p) \, =\, [2]\, A(p)\,.
\ee
According to \cite{HIOPT} the general $SL(n)$-type dynamical $R$-matrix \cite{I2}
can be obtained from (\ref{A1}) by either an analog of Drinfeld's twist
\cite{D2} (see Lemma 3.1 of \cite{HIOPT}) or by a canonical transformation
$p_i \to p_i +c_i\,$ where $c_i\,$ are constants (numbers) such that
$\sum_{i=1}^n c_i = 0\,$. The interpretation of the eigenvalues $p_i\,$ of
${\hat p}_i\,$ as (shifted) weights (of the corresponding representations of
$U_q\,$) allows to dispose of the second freedom.

Inserting (\ref{A1}) into the exchange relations (\ref{Rpaa=aaR}) allows
to present the latter in the following explicit form:
\setcounter{equation}{27}
\renewcommand{\theequation}{\thesection.\arabic{equation}}
\be
\lb{aa1}
[a^i_\a , a^j_\a ] = 0\,,\quad
a^i_\a a^i_\b = q^{{\epsilon}_{\a\b}} a^i_\b a^i_\a
\ee
\be
\lb{aa2}
[p_{ij}-1]\, a^j_\a a^i_\b = [p_{ij}] a^i_\b a^j_\a -
q^{{\epsilon}_{\b\a}p_{ij}} a^i_\a a^j_\b\ \ {\rm for}\ \ \a\ne\b\ \
{\rm and}\ i\ne j,
\ee
where $q^{\epsilon_{\alpha\beta}}$ is defined in (1.28b).

There is, finally, a relation of order $n$ for $a^i_\a\,$, derived from
the following basic property of the {\em quantum determinant}:
\be
\lb{qdet}
\det (a)\, = \, {1\over{[n]!}}\,
\edo \ai{1} \dots \ai{n} \Eup\equiv
{1\over{[n]!}}\, {\varepsilon}_{\scriptscriptstyle i_1 \dots i_n}
a^{i_1}_{\a_1} \dots a^{i_n}_{\a_n}\,
{\cal E}^{\scriptscriptstyle \a_1 \dots \a_n}
\ee
where ${\cal E}^{\scriptscriptstyle \a_1 \dots \a_n}\,$
is given by (\ref{e}) while
${\varepsilon}_{\scriptscriptstyle i_1 \dots i_n} \,$ is the
dynamical Levi-Civita tensor with lower indices (which can be
consistently chosen to be equal to the undeformed one \cite{HIOPT},
a convention which we assume throughout this paper), normalized by
${\varepsilon}_{\scriptscriptstyle n\dots 1} = 1 \,$. The ratio
$\det (a) \left( \prod_{i<j} [p_{ij}] \right)^{-1}\,$ belongs to the
centre of the quantum matrix algebra
${\cal A} = {\cal A} (\Rp , \R )\,$ (see Corollary 5.1 of
Proposition 5.2 of \cite{HIOPT}).
It is, therefore, legitimate to normalize the quantum determinant setting
\be
\lb{qdnorm}
\det (a)\, =\,
\prod_{i<j}\, [p_{ij}]
\equiv \, {\cal D} (p)\,.
\ee
It is proportional (with a positive $p$-independent factor) to
the quantum dimension (\ref{qdim}).

\vspace{5mm}

\noindent
{\em Remark 2.2~} The results of this section are
clearly applicable if the determinant ${\cal D}(p)\,$ does not
vanish (i.e., either for generic $q\,$ or, if $q\,$ is given by
(\ref{0.1}), for $p_{1n}<h\;$). As noted in the introduction, the
notion of a CVO and the splitting (\ref{vIRF}) may well require a
modification if this condition is violated.

\vspace{5mm}

To sum up: {\em the intertwining quantum matrix algebra ${\cal
A}\,$ is generated by the $n^2\,$ elements $a^i_\a\,$ and the
field ${\Bbb Q} (q, q^{p_i})\,$ of rational functions of the
commuting variables $q^{p_i}\,$ whose product is $1\,$, subject
to the exchange relations (\ref{qa}) and (\ref{Rpaa=aaR}) and the
determinant condition (\ref{qdnorm}).}

The centralizer of $q^{p_i}\,$ in
${\cal A}\,$ (i.e., the maximal subalgebra of ${\cal A}\,$
commuting with all $q^{p_i}\;$) is spanned by
the QUEA $U_q\,$ over the field ${\Bbb Q} (q, q^{p_i})\,$ and
$a^i_\a\,$ obey the $U_q\,$ covariance relations (2.10). The
expressions for the $U_q\,$ generators in terms of $n$-linear
products of $a^i_\a\,$ are worked out for $n=2\,$ and $n=3\,$ in
Appendix A.

We shall use in what follows the intertwining properties of the product
$a_1\dots a_n\,$ (see Proposition 5.1 of \cite{HIOPT}):
\setcounter{equation}{0}
\renewcommand{\theequation}{\thesection.32\alph{equation}}
\be
\lb{int1}
\edo\,a_1\dots a_n\, =\, {\cal D}(p)\,\Edo
\ee
or, in components,
\be
\lb{int2}
{\varepsilon}_{\scriptscriptstyle i_1 \dots i_n} \,
a^{i_1}_{\a_1}\dots
a^{i_n}_{\a_n}\, = \, {\cal D}(p)\,
{\cal E}_{\scriptscriptstyle \a_1 \dots \a_n}\,;
\ee
\setcounter{equation}{32}
\renewcommand{\theequation}{\thesection.\arabic{equation}}
\be
\lb{int3}
a_1\dots a_n \,\Eup \,=\, \eupp\, {\cal D}(p)\,.
\ee
Here ${\varepsilon}(p)\,$ is the dynamical
Levi-Civita tensor with upper indices given by
\be
\lb{dLC}
{\varepsilon}^{\scriptscriptstyle i_1 \dots i_n}(p)\,
=\, (-1)^{\ell(\sigma )}\,\prod_{1\le\mu <\nu\le n}
{ {[p_{i_\mu i_\nu} -1]}\over{[p_{i_\mu i_\nu}]} }\,,
\ee
$\ell(\sigma )\,$ standing again for the length of the permutation
$\sigma = {n_{~},\dots ,~1_{~}\choose i_1,\dots ,~i_n}\,$.

\vspace{5mm}
\noindent
{\em Remark 2.3~} Selfconsistency of
(\ref{1.17}) requires that ${\rm det} (a) = {\rm det} (a M)\,.$
Indeed, the non-commutativity of $q^{{p}_j}\,$ and $a^i\,,$ see
Eq.(\ref{qa}), exactly compensates the factors $q^{1-{1\over
n}}\,$ when computing the determinant of $a M\,$ (cf. (2.11a),
(2.11b)); we have \be q^{2{p}_n -1+{1\over n}}\, a^n_{\a_1}\,
q^{2{p}_{n-1} -1+{1\over n}}\, a^{n-1}_{\a_2}\,\ldots q^{2{p}_1
-1+{1\over n} }\, a^1_{\a_n} \, =\, a^n_{\a_1}
a^{n-1}_{\a_2}\ldots a^1_{\a_n} \ee since \be q^{{2\over n}
(1+2+\dots + n-1) - n + 1}\, =\, 1\,. \ee

\vspace{5mm}

An important consequence of the ice property (\ref{DDR}) (valid for both
$\R\;$ and $\Rp\;$) is the existence of subalgebras of ${\cal A}\,$ with
similar properties.

Let
$$I=\{i_1,i_2,\dots ,i_m\}\,,\quad 1\leq i_1<i_2<\dots <i_m\leq n$$
and
$$\Gamma=\{\alpha_1, \alpha_2,\dots ,\alpha_m\}\,,\quad 1\leq
\alpha_1<\alpha_2<\dots <\alpha_m\leq n\,$$
be two ordered sets of $m\,$ integers
($1\leq m\leq n\,$). Let $A_{1\,m}|_{\Gamma}$ be the restriction of the
$q$-antisymmetrizer $(A_{1\,m})^{\alpha_1 \alpha_2 \dots \alpha_m}_{\beta_1
\beta_2 \dots \beta_m}\,,\ \,\alpha_k, \beta_k \in \{ 1,2,\dots ,n\}$ (for its
definition see (1.30)) to a subset of indices $\alpha_k, \beta_k
\in\Gamma\,.$ Then $\rank A_{1\,m}|_{\Gamma}=1\,$ and one can define the
corresponding restricted Levi-Civita tensors satisfying
\be
\lb{LC-restr}
A_{1\,m}|_{\Gamma}=  {{\cal E}|_{\Gamma}}^{|1  \dots
m{\cal i} }\,{{\cal E}|_{\Gamma}}_{{\cal h}1  \dots m| }\,,\qquad
{{\cal E}|_{\Gamma}}_{{\cal h}1  \dots m| }
{{\cal E}|_{\Gamma}}^{|1  \dots m{\cal i}}=[m]!\ .
\ee
In the same way one defines restricted dynamical Levi-Civita
tensors
$${{\varepsilon}|_{I}}^{|1 \dots m{\cal i} }(p)\quad {\rm
and}\quad {{\varepsilon}|_{I}}_{{\cal h}1 \dots m| }(p)
$$
for the subset $I\subset \{1,2,\dots ,n\}$ (the last one of these does not
actually depend on $p$ and coincides with the classical Levi-Civita
tensor).

Consider the subalgebra ${\cal A}(I,\Gamma)\subset {\cal A}$ generated by
${\Bbb Q}(q, q^{p_{ij}})$, $i,j\in I$ and the elements of the
submatrix $a|_{I,\Gamma}\, : = {\|}a{\|}^{i\in I}_{\alpha\in\Gamma}$
of the quantum matrix $a\,$.

\vspace{5mm}

\noindent
{\bf Proposition 2.4}~ {\em The normalized minor \be
\lb{D} \Delta_{I,\Gamma}(a) := {\det(a|_{I,\Gamma})\over {\cal
D}_I(p)} := {1\over [m]! {\cal D}_I(p)}
{{\varepsilon}|_{I}}_{{\cal h}1  \dots m| } (a_1 a_2 \dots
a_m)|_{I,\Gamma}\, {{\cal E}|_{\Gamma}}^{|1  \dots m{\cal i} }\ ,
\ee where \be \lb{D2} {\cal D}_I(p):= \prod_{i<j;~ i,j\in I}
[p_{ij}] \ee belongs to the centre of ${\cal A}(I,\Gamma)$. }
\vspace{5mm}

The statement follows from the observation that relations
(\ref{int1}--\ref{int3}) and (\ref{dLC})
are valid for the restricted quantities
${\cal E}|_{\Gamma}$, ${\varepsilon}|_I$, ${\cal D}_I(p)$ and
$a|_{I,\Gamma}$.
\eod

\vspace{5mm}

Using restricted analogs of the relations (\ref{int3}),
(\ref{dLC}),
we can derive alternative expressions for the normalized minors:
\be
\lb{D3}
\Delta_{I,\Gamma}(a) =
{1\over {\cal D}_I^+(p)} a^{i_1}_{\alpha_1} a^{i_2}_{\alpha_2}\dots
a^{i_m}_{\alpha_m} {\cal E}|_{\Gamma}^{\alpha_1\dots \alpha_m}\ ,
\ee
where
\be
\lb{D4}
{\cal D}^+_I(p) := \prod_{i<j;~ i,j\in I}[p_{ij} +1]\ ,
\ee
the indices $i_k \in I$ are in descendant order,
$i_1> i_2 >\dots >i_m\,,$
and the indices $\alpha_k\in \Gamma$ are summed over.

\vspace{5mm}

\section{The Fock space representation of ${\cal A}\,$.
The ideal ${\cal I}_h\,$ for $q^h = -1\,$.}
\setcounter{equation}{0}
\renewcommand{\theequation}{\thesection.1\alph{equation}}

\vspace{5mm}

\subsection{The Fock space ${\cal F} ({\cal A})\,$ (the case of
generic $q\;$).}

\medskip

The "Fock space" representation of the quantum matrix algebra ${\cal
A}\,$ was anticipated in Eq.(2.7) and the subsequent discussion
of Young tableaux. We define ${\cal F}\,$ and its dual ${\cal F}'\,$
as cyclic ${\cal A}\;$
modules with one dimensional $U_q$-invariant subspaces of
multiples of (non-zero) {\em bra and ket vacuum vectors}
${\cal h} 0 |\,$
and $| 0 {\cal i}\,$
such that ${\cal h} 0 | {\cal A} = {\cal F}'\,,\
{\cal A} |0 {\cal i} = {\cal F}\,$
satisfying
\ba
&& a^i_\a | 0 {\cal i} = 0
\quad {\rm for}\quad i>1,
\quad {\cal h} 0 | a^j_\a = 0
\quad {\rm for}\quad j<n,\\
&& q^{p_{ij}} | 0 {\cal i} = q^{j-i} | 0 {\cal i}\,,\quad
{\cal h} 0 |\; q^{p_{ij}} = q^{j-i} {\cal h} 0 |\,,\\
&& (X - \varepsilon (X) ) | 0 {\cal i}
= 0 = {\cal h} 0 | (X - \varepsilon (X) )
\ea
for any $X\in U_q\,$ (with $\varepsilon (X)\,$ the counit). The duality between
${\cal F}\,$ and ${\cal F}'\,$ is established by a bilinear
pairing ${\cal h}\; .\; |\; .\; {\cal i}\,$ such that
\setcounter{equation}{1}
\renewcommand{\theequation}{\thesection.\arabic{equation}}
\be
\lb{dual}
{\cal h} 0 | 0 {\cal i} = 1\,,\quad
{\cal h} \Phi | A | \Psi {\cal i} =
{\cal h} \Psi | A' | \Phi {\cal i}
\ee
where $A \to A'\,$ is a linear antiinvolution ({\em
transposition}) of ${\cal A}\,$
defined for generic $q\,$ by
\be
\lb{'}
{\cal D}_i(p) (a^i_\a)' = {\tilde a}^\a_i := {1\over{[n-1]!}}\;
{\cal E}^{\scriptscriptstyle \a\a_1 \dots \a_{n-1}}\;
{\varepsilon}_{\scriptscriptstyle i i_1 \dots i_{n-1}}\;
a^{i_1}_{\a_1}\dots a^{i_{n-1}}_{\a_{n-1}}\;,\ \ (q^{p_i})' = q^{p_i}\,,
\ee
${\cal D}_i(p)\,$ standing for the product
\be
\lb{minor}
{\cal D}_i(p) = \prod_{j<l,\, j\ne i\ne l} [p_{jl}]\quad \left(
 \Rightarrow\ [{\cal D}_i(p), a^i_\a ] = 0 = [{\cal D}_i(p), {\tilde a}^\a_i ] \right)\,.
\ee
We verify in Appendix B the involutivity
property, $A'' = A\,,$ of (\ref{'}) for $n=3\,.$
Eq. (\ref{'}) implies the following formulae
for the transposed of the Chevalley generators of $U_q\,$:
\be
\lb{Uq'}
E_i\;' = F_i\; q^{H_i-1}\,,\quad F_i\;' = q^{1-H_i} E_i\,,\quad (q^{H_i})' =
q^{H_i}\,.
\ee

The main result of this section is the proof of the statement that
for generic $q\,$ ($q\,$ not a root of
unity) ${\cal F}\,$ is a model space for $U_q\;$: each finite dimensional IR of
$U_q\,$ is encountered in ${\cal F}\,$ with multiplicity one.

\vspace{5mm}

\noindent
{\bf Lemma 3.1}~{\em For generic $q\,$ the space ${\cal F}\,$ is spanned
by antinormal ordered
polynomials applied to the vacuum vector:
\be
\lb{onvac}
P_{m_{n-1}}(a^{n-1}_\a )\dots P_{m_1}(a^1_\a ) \; | 0{\cal i}\quad
{\rm with}\quad m_1\ge m_2\ge\dots\ge m_{n-1}\,.
\ee
Here $P_{m_i}(a^i_\a )\,$ is a homogeneous polynomial
of degree $m_i\,$ in $a^i_1 ,\dots , a^i_n\,.$
}

\vspace{5mm}

\noindent
{\em Proof}~~We shall first prove
the weaker statement that ${\cal F}\,$ is
spanned by vectors of the type
$P_{m_{n}}(a^{n}_\a )\dots P_{m_1}(a^1_\a ) \; | 0{\cal i}\,$
(without restrictions on the
nonnegative integers $m_1, \dots ,m_n\,$). It follows from the exchange relations
(\ref{aa2}) for
$i>j\,$ and from the observation that $[p_{jl}+1]\ne 0\,$ for generic $q\,$ and $j<l\,$ in
view of (3.1b).

Next we note that if $m_{j-1} = 0\,$ but $m_j >0\,$ for some $j>1\,,$ the
resulting vector vanishes. Indeed, we can use in this case repeatedly (\ref{aa2})
for $i<j-1\,$ to move an $a^j_\a\,$ until it hits the vacuum giving zero according to (3.1a).

If all $m_i >0\,, \ i=1,\dots , n\,,$ we move a factor $a^i_{\a_i}\,$ of each monomial
to the right to get rid of an $n$-tuple of $a^i_{\a_i}\,$ since
\be
\lb{navac}
a^n_{\a_n} \dots a^1_{\a_1} |0{\cal i} = [n-1]!
{\cal E}_{\scriptscriptstyle \a_n\dots \a_1} |0{\cal i}\;;
\ee
here we have used once more (3.1a), and also (2.32) and (3.1b).
Repeating this procedure $m_n\,$ times, we obtain  an expression of the type
(\ref{onvac}) (or zero, if $m_n >\, {\rm min}\,(m_1,\dots ,m_{n-1})$ ).

To prove the inequalities $m_i \ge m_{i+1}\,$ we can reduce the problem (by the same procedure
of moving whenever possible $a^i_\a\,$ to the right) to the statement that any expression of
the type
$a^{i+1}_{\b_1} a^{i+1}_{\b_2} a^i_{\a_i} \dots a^1_{\a_1} |0 {\cal i}\,$ vanishes.
We shall display the argument for a special case proving that
\be
\lb{3321}
a^3_\a a^3_\b a^2_2 a^1_1 | 0 {\cal i} = 0\quad {\rm for}\ n\ge 3\,.
\ee
This is a simple consequence of (\ref{aa1}), (\ref{aa2})
and (3.1a) if either $\a\,$ or $\b\,$ is
$1\,$ or
$2\,$. We can hence write, using (2.10b),
\be
\lb{F23321}
0 = F_2 a^3_2 a^3_3 a^2_2 a^1_1 | 0 {\cal i} =
(a^3_3 )^2 a^2_2 a^1_1 | 0 {\cal i} +
a^3_2 a^3_3 a^2_3 a^1_1 | 0 {\cal i} =
(a^3_3 )^2 a^2_2 a^1_1 | 0 {\cal i}\,.
\ee
By repeated application of $F_i\,$ (with $i\ge 3\,$ for $n\ge 4\,$) exploiting the $U_q\,$
invariance of the vacuum (3.1c), we thus complete the proof of (\ref{3321}) and hence, of
Lemma 3.1.
\eod

\vspace{5mm}

\noindent
{\bf Corollary}~{\em It follows from Lemma 3.1 that the space
${\cal F}\,$ splits into a direct sum of weight spaces ${\cal
F}_p\,$ spanned by vectors of type (\ref{onvac}) with fixed
degrees of homogeneity $m_1,\dots , m_{n-1}\,$:
\be
\lb{Fspace}
{\cal F} = \oplus_p {\cal F}_p\,,\quad p_{ij} = m_i-m_j+j-i\, (\ge
j-i\ \ {\rm for}\ \ i<j)\,,
\ee
each subspace ${\cal F}_p\,$ being characterized by (\ref{q^p}).}

\vspace{5mm}

In order to exhibit the $U_q\,$ properties of ${\cal F}_p\,$ we
shall introduce the {\em highest} and {\em lowest weight vectors}
(HWV and LWV)
$$|\l_1  \dots \l_{n-1} {\cal i}\qquad {\rm and} \qquad
|-\l_{n-1}\,\dots\,-\l_1 {\cal i}\,,$$
obeying
\be
\lb{Cart}
(q^{H_i}-q^{\lambda_i})| \l_1\dots \l_{n-1} {\cal i} = 0 =
(q^{H_i} - q^{-\lambda_{n-i}} )| -\l_{n-1}\,\dots\, -\l_1  {\cal i}
\ee
for $\l_i = m_i - m_{i+1} = p_{i\; i+1} - 1\,,\quad 1\le i\le n-1\,.$

\vspace{5mm}

\noindent

{\bf Lemma 3.2}~{\em Each ${\cal F}_p\,$ contains a unique (up to
normalization) HWV and a unique LWV satisfying (\ref{Cart}). They
can be written in either of the following three equivalent forms:
\ba
\lb{monom}
&&| \lambda_1 \dots \lambda_{n-1} {\cal i} = \\
&&= (\Delta^{n-1\;1}_{n-1\;1})^{\lambda_{n-1}}
(\Delta^{n-2\;1}_{n-2\;1})^{\lambda_{n-2}}
\dots (\Delta^{2\;1}_{2\;1})^{\lambda_2}(a^1_1)^{\lambda_1}
| 0 {\cal i} =\nonumber\\
&&= (a^1_1)^{\lambda_1}
(\Delta^{2\;1}_{2\;1})^{\lambda_2}\dots (\Delta^{n-1\;1}_{n-1\;
1})^{\lambda_{n-1}} |0{\cal i}
\sim\nonumber\\
&&\sim (a^{n-1}_{n-1})^{\lambda_{n-1}}
(a^{n-2}_{n-2})^{\lambda_{n-2}+\lambda_{n-1}}
\dots (a^1_1)^{\lambda_1 +\dots +\lambda_{n-1}}
|0{\cal i} \,,\nonumber
\ea
\ba
\lb{LWV}
&&| - \lambda_{n-1}\,\dots \,-\lambda_1 {\cal i} =\\
&&=
(\Delta^{n-1\;1}_{n\;2})^{\lambda_{n-1}} (\Delta^{n-2\;1}_{n\;
3})^{\lambda_{n-2}}
\dots (\Delta^{2\;1}_{n\;n-1})^{\lambda_2}(a^1_n)^{\lambda_1}
| 0 {\cal i} = \nonumber\\
&&= (a^1_n)^{\lambda_1}
(\Delta^{2\;1}_{n\;n-1})^{\lambda_2}\dots (\Delta^{n-1\;1}_{n\;
2})^{\lambda_{n-1}} |0{\cal i}\sim\nonumber\\
&&\sim (a^{n-1}_2)^{\lambda_{n-1}}
(a^{n-2}_3)^{\lambda_{n-2}+\lambda_{n-1}}\dots (a^1_n)^{\lambda_1
+\dots +\lambda_{n-1}} |0{\cal i} \,.;\nonumber
\ea
here $\Delta^{i\;1}_{i\;1}\,$
and $\Delta^{i\;1}_{n\; n-i+1}\,$
are normalized minors of the type (\ref{D3}):
\be
\lb{Del}
\Delta^{i\;1}_{i\;1} = \Delta_{I_i, \Gamma_i}(a) =
{1\over {\cal D}^+_{I_i}(p)} \, a^i_{\alpha_1}\dots
a^1_{\alpha_i}{\cal E}|_{\Gamma_i}^{\alpha_1\dots \alpha_i}
\ee
for $I_i := \{1,2,\dots ,i\} =: \Gamma_i\,,$ and
\be
\lb{Dell}
\Delta^{i\; 1}_{n\; n-i+1} \equiv \Delta_{I_i, \Gamma^i_n}(a) =
{1\over {\cal D}^+_{I_i}(p)}\, a^i_{\alpha_1}\dots
a^1_{\alpha_i}{\cal E}|_{\Gamma^i_n}^{\alpha_1\dots \alpha_i}
\ee
where $\Gamma^i_n :=\{n-i+1,n-i+2,\dots ,n\}\,.$
}

\vspace{5mm}

\noindent
{\em Proof}~~We shall prove the uniqueness of the HWV
by reducing an arbitrary eigenvector of
$q^{H_i}\,$ of eigenvalue $q^{\lambda_i}\,,\ 1\le i\le n-1\,,$
to the form of the second equation (\ref{monom}).
To this end we again apply the argument in the proof of Lemma 3.1.
Let $k\, (\le n-1)\;$ be the maximal numeral for which $\lambda_k >0\,.$
By repeated application
of the exchange relations (\ref{aa2}) we can arrange each $k$-tuple $a^k_{\a_1}
\dots a^1_{\a_k}\,$ to hit a vector $| v {\cal i}\,$ such that
$(p_{i i+1}-1 ) | v {\cal i} = 0\ {\rm for}\ i<k\,$. (Observe that
all vectors of the type $|v_k{\cal i}= (\Delta^{k\;
1}_{k\;1})^{\lambda_k}\dots (\Delta^{n-1\;1}_{n-1\;
1})^{\lambda_{n-1}} |0{\cal i}\,,$ for various choices of the
non-negative integers $\l_k ,\dots , \l_{n-1}\,,$ have this
property.) Noting then that $a^{i+1}_\a | v {\cal i} = 0\,$
whenever $(p_{i i+1}-1 ) | v {\cal i} = 0\,$ and using once more
Eq.(\ref{aa2}) we find
\be
\lb{skews}
( p_{i i+1}-1 ) | v {\cal i} = 0 \quad\Leftrightarrow\quad
(a^{i+1}_\b a^i_\a +
q^{\epsilon_{\a\b}} a^{i+1}_\a a^i_\b ) | v {\cal i} = 0
\ee
which implies that we can substitute the product $a^{i+1}_\a a^i_\b\,$ (acting on
such a vector) by its antisymmetrized expression:
\be
\lb{antis}
a^{i+1}_\a a^i_\b
| v {\cal i} =
{1\over{[2]}}
(q^{\epsilon_{\b\a}}
a^{i+1}_\a a^i_\b - a^{i+1}_\b a^i_\a )
| v {\cal i} \quad (\;{\rm for}\ (p_{i i+1}-1) | v{\cal i} = 0\, )\;.
\ee
Such successive antisymmetrizations will
give rise to the minor $\Delta^{k 1}_{k
1}\,$ yielding eventually the second
expression  (\ref{monom}) for the HWV.

To complete the proof of Lemma 3.2, it remains to prove the
first equalities in (\ref{monom}) and (\ref{LWV}). The
commutativity of all factors $\Delta^{i\;1}_{i\;1}\,,\ 1\le i\le
n-1\ (\Delta^{1\;1}_{1\;1} \equiv a^1_1\,)\,$ follows from
Proposition 2.4 which implies
\be
\lb{aD} [ a^i_\a\,, \,
\Delta^{k\;1}_{k\;1} ] = 0\quad {\rm for}\quad 1\le \a ,i \le k\,.
\ee
In order to compute the proportionality factors between the
second and the third expressions in (\ref{monom}), (\ref{LWV})
one may use the general exchange relation
\be
\lb{genex}
[p_{ij}-1](a^j_\a)^ma^i_\b = [p_{ij}+m-1]a^i_\b(a^j_\a)^m -
q^{\epsilon_{\b\a}(p_{ij}+m-1)} [m] (a^j_\a)^{m-1}a^i_\a a^j_\b
\ee
(valid for $i\ne j\,$ and $\a\ne\b\,$) which follows from
(\ref{aa2}).

\vspace{5mm}

Lemmas 3.1 and 3.2 yield the main result of this section.

\vspace{5mm}

\noindent
{\bf Proposition 3.3~} {\em The space ${\cal F}\,$ is (for generic
$q\;$) a model space of $U_q\,.$}
\eod

\vspace{5mm}

We proceed to defining the $U_q\,$ symmetry of a Young tableau $Y\,$.
A $U_q\,$ tensor $T_{\a_1\dots \a_s}\,$
is said to be {\em $q$-symmetric} if for any pair of
adjacent indices $\a \;\b\,$ we have
\be
\lb{qsym}
T_{\dots{\a}'{\b}'\dots} A^{{\a}'{\b}'}_{\a\b} =
0\quad\Leftrightarrow\quad T_{\dots\a\b\dots}
= q^{\epsilon_{\a\b}} T_{\dots\b\a\dots}
\ee
where $q^{\epsilon_{\a\b}}\,$ is defined in (1.28b). A tensor
$F_{\a_1\dots \a_s}\,$ is {\em $q$-skewsymmetric} if it is an eigenvector of the
antisymmetrizer (1.28b):
\be
\lb{qskew}
F_{\dots{\a}'{\b}'\dots} A^{{\a}'{\b}'}_{\a\b} =
[2] F_{\dots{\a}{\b}\dots}\quad\Leftrightarrow\quad
F_{\dots{\a}{\b}\dots} = - q^{\epsilon_{\b\a}}  F_{\dots\b\a\dots}\,.
\ee
{\em A $U_q\,$ tensor of
$\lambda_1+ 2\lambda_2+\dots +(n-1)\lambda_{n-1}\,$
indices has the $q$-symmetry of a
Young tableau $Y = Y_{[\lambda_1 ,\dots ,\lambda_{n-1}]}\,$
(where $\lambda_i\,$ stands for
the number of columns of height $i\,$)
if it is first $q$-symmetrized in the indices of each
row and then $q$-antisymmetrized along the columns.}

The $q$-symmetry of a
tensor associated with a Young tableau allows to choose as independent
components an ordered set of values of the indices $\a,\b\,$ that
monotonically increase
along rows and strictly increase down the columns (as in the undeformed case - see
\cite{Ful}). Counting such labeled tableaux of a fixed type $Y\,$ allows to reproduce the
dimension $d_1(p)\,$ of the space ${\cal F}_p\,.$

\vspace{5mm}

\subsection{Canonical basis. Inner product.}

\medskip

We shall introduce a {\em canonical basis} in
the $U_q\,$ modules ${\cal F}_p\,$
in the simplest cases of $n = 2,3\,$ preparing the ground
for the computation of inner products in ${\cal F}_p\,$ for such
low values of $n\,$.

We shall follow Lusztig \cite{L}
for a general definition of a canonical basis. It is, to
begin with, a {\em basis of weight vectors},
a property which determines it (up to normalization) for
$n=2\,.$ We shall set in this case
\be
\lb{can2}
|p,  m{\cal i} = (a^1_1)^m (a^1_2)^{p-1-m}\; |0{\cal i}\,,\quad
0\le m\le p-1\ \ (p\equiv p_{12})\,.
\ee
Introducing (following \cite{L}) the operators
\be
\lb{[m]}
E^{[m]} = {1\over{[m]!}} E^m\,,\quad F^{[m]} = {1\over{[m]!}} F^m
\ee
we can relate $|p, m {\cal i}\,$ to the HWV and LWV in ${\cal F}_p\,$:
\be
\lb{|m>}
F^{[p-1-m]} |p, p-1 {\cal i} = \left[{{p-1}\atop{m}}\right]
|p, m{\cal i} = E^{[m]} |p, 0{\cal i}\,.
\ee

The situation for $n=3\,$ can still be handled more or less explicitly.
A basis in ${\cal F}_p\,$ is constructed in that case by applying
Lusztig's canonical basis \cite{L} in either of the two conjugate
Hopf subalgebras of raising or lowering  operators
\be
\lb{XXX}
X_1^{[m]}X_2^{[\ell]}X_1^{[k]}\quad {\rm and}\quad
X_2^{[k]}X_1^{[\ell]}X_2^{[m]}\quad {\rm for}\quad
X= E\ {\rm or}\ F\,,\ \ell \ge
k+m\,,
\ee
the $U_q\,$ Serre relations implying
\be
\lb{X3=X3}
X_1^{[m]}X_2^{[k+m]}X_1^{[k]}\,=\,
X_2^{[k]}X_1^{[k+m]}X_2^{[m]}\,,
\ee
to the lowest or to the highest weight vector, respectively,
\be
\lb{lw}
E_1^{[m]}E_2^{[\ell ]}E_1^{[k]}\,
|-\lambda_2\, - \lambda_1 {\cal i}\,, \quad
E_2^{[k]}E_1^{[\ell ]}E_2^{[m]}\,
|-\lambda_2\, - \lambda_1 {\cal i}\,,
\ee
\be
\lb{hw}
F_1^{[m]}F_2^{[\ell ]}F_1^{[k]}\,
|\lambda_1  \lambda_2 {\cal i}\,, \quad
F_2^{[k]}F_1^{[\ell ]}F_2^{[m]}\,
|\lambda_1  \lambda_2 {\cal i}\,,\quad 0\le
k+m\le \ell \le \lambda_1 +\lambda_2
\ee
where we are setting
\ba
&&|\lambda_1  \lambda_2 {\cal i}
= (a^1_1)^{\lambda_1} (q{a^3_3}')^{\lambda_2}
|0{\cal i}\,,\\
&&|-\lambda_2\, - \lambda_1 {\cal i} =
(a^1_3 )^{\lambda_1} (\bq{a^3_1}')^{\lambda_2}
|0{\cal i}\,.
\ea
(These expressions differ by an overall power of $q\,$ from (\ref{monom})
and
(\ref{LWV}).)

\vspace{5mm}

\noindent
{\bf Lemma 3.4}~{\em The action of $F_i^{[m]}\ \,(E_i^{[m]} )\,,\
m\in{\Bbb N}$ on a HWV (LWV) is given by}
\setcounter{equation}{0}
\renewcommand{\theequation}{\thesection.31\alph{equation}}
\ba
F_1^{[m]} |\lambda_1  \lambda_2 {\cal i} &=&
\left[{\lambda_1\atop m} \right] (a^1_1 )^{\lambda_1
-m}
(a^1_2)^m
(q{a^3_3}')^{\lambda_2} |0{\cal i}\,,\\
F_2^{[m]} |\lambda_1  \lambda_2 {\cal i} &=&
\left[{\lambda_2\atop m} \right] (a^1_1 )^{\lambda_1}
(q{a^3_3}')^{\lambda_2 -m} (-{a^3_2}')^m |0{\cal i}\,,\nonumber\\
E_1^{[m]} |-\lambda_2\, - \lambda_1 {\cal i} &=&
\left[{\lambda_2\atop m}\right] (a^1_3 )^{\lambda_1} (-{a^3_2}')^m
(\bq{a^3_1}')^{\lambda_2 -m} |0{\cal i}\,,\\
E_2^{[m]} |-\lambda_2\, - \lambda_1 {\cal i} &=&
\left[{\lambda_1\atop m}\right] (a^1_2)^m (a^1_3 )^{\lambda_1 -m}
(\bq{a^3_1}')^{\lambda_2} |0{\cal i}\,.\nonumber
\ea

\vspace{5mm}

The {\em proof} uses (2.10), (\ref{aa1}) and the relations
\setcounter{equation}{31}
\renewcommand{\theequation}{\thesection.\arabic{equation}}
\be
\lb{qcomanti}
{a^3_2}' {a^3_3}' = q\; {a^3_3}' {a^3_2}'
\,,\quad
{a^3_1}' {a^3_2}' = q\; {a^3_2}' {a^3_1}'
\ee
obtained by transposing the second equality in (\ref{aa1}) for $i=3\,$.

\vspace{5mm}

We shall turn now to the computation of the $U_q\,$ invariant form.

\vspace{5mm}

\noindent
{\bf Conjecture 3.5}~{\em The scalar square of the HWV
(\ref{monom}) and the LWV (\ref{LWV}) of $U_q\,$ is given by}
\be
\lb{scsq}
{\cal h} \lambda_1\dots \lambda_{n-1}
| \lambda_1\dots \lambda_{n-1} {\cal i} =
\prod_{i<j} [p_{ij}-1]! =
{\cal h} -\lambda_{n-1}\dots -\lambda_1
| -\lambda_{n-1}\dots -\lambda_1 {\cal i}\,.
\ee

\vspace{5mm}

For $n=2\,$ the result is a straightforward consequence of Eqs.(\ref{can2}) and
(\ref{A.11}) (of Appendix A). For $n=3\,$ Eq.(\ref{scsq}) reads
\be
\lb{scsq3}
{\cal h} \lambda_1 \lambda_2 | \lambda_1 \lambda_2 {\cal i} =
[\lambda_1 ]![\lambda_2 ]! [\lambda_1 + \lambda_2 +1]! =
{\cal h} -\lambda_2\, -\lambda_1 | -\lambda_2\, -\lambda_1 {\cal i}
\ee
which is proven in Appendix C.
We conjecture that the argument can be extended to prove (\ref{scsq}) for any $n\ge
2\,.$

\vspace{5mm}

For $n=2\,$ we can also write the inner products
of any two vectors of the canonical basis \cite{FHT3}:
\be
\lb{scpr2}
{\cal h} p, m | p' , m'{\cal i} = \d_{p p'} \d_{m m'}
{\bq}^{m(p-1-m)} [m]! [p-1-m]!\,.
\ee

\vspace{5mm}

\subsection{The case of $q\,$ a root of unity.
Subspace of zero norm vectors. Ideals in ${\cal A\,}$.}

\medskip

In order to extend our results to the study
of a WZNW model, we have to
describe the structure of the $U_q\,$ modules
${\cal F}_p\,$ for $q\,$ a root of unity, (\ref{0.1}).
Here ${\cal F}_p\,$ is, by definition, the space spanned by
vectors of type (\ref{onvac}) (albeit the proof of Lemma 3.1 does not
apply to this case).
We start by recalling the situation for $n=2\,$ (see \cite{FHT3,
DT, goslar}).

\medskip

The relations
\be
\lb{action2}
E |p, m{\cal i} = [p-m-1] |p, m+1{\cal i}\,,\quad
F |p, m{\cal i} = [m] |p, m-1 {\cal i}
\ee
show that for $p\le h\,$ the $U_q\,$ module ${\cal F}_p\,$
admits a single HWV and LWV and
is, hence, irreducible. For $p>h\,$ the situation changes.

\vspace{5mm}

\noindent
{\bf Proposition 3.6}~{\em
For $h<p<2h\,$ and $q\,$ given by (\ref{0.1})
the module ${\cal F}_p\,$ is indecomposable.
It has two $U_q(sl_2)\,$ invariant subspaces with no invariant
complement:
\ba
\lb{invsub}
{\cal I}^+_{p,h} &=&{\rm Span}\,\{|p, m{\cal i}\;,\ h\le m\le
p-1\;\}\;,\nonumber
\\ \\
{\cal I}^-_{p,h} &=&{\rm Span}\,\{|p, m{\cal i}\;,\ 0\le m\le
p-1-h\;\}\;.\nonumber
\ea
It contains a second pair of singular vectors: the LWV
$|p, h{\cal i}\,$ and the HWV $|p, p-1-h{\cal i}\,.$
The vector $|p, p-h{\cal i}\;$ is {\em cosingular}, -- i.e., it
cannot be written in the form $E |v{\cal i}\,$ with $v\in {\cal
F}(p)\,;$ similarly, the vector $|p, h-1{\cal i}\,$ cannot be
presented as $F |v{\cal i}\,.$ }

\vspace{5mm}

The statement follows from (\ref{action2}) and from
the fact that
\be
\lb{fact}
F |p, p-h{\cal i} = [p-h] |p, p-h-1{\cal i} \ne 0\ne E |p, h-1{\cal i}
\ee
so that
the invariant subspace ${\cal I}^+_{p,h}\oplus {\cal I}^-_{p,h}\,$
indeed has no invariant complement in ${\cal F}_p\,$.

\vspace{5mm}

The factor space
\be
\lb{fac}
{\tilde{\cal F}}_p\, = \,
{\cal F}_p / ( {\cal I}^+_{p,h}\oplus {\cal I}^-_{p,h} )\ \quad
(h<p<2h)
\ee
carries an IR of $U_q(sl_2)$ of weight ${\tilde p} = 2h-p\,$ (cf.
(\ref{singvect})).

\vspace{5mm}

The inner product (\ref{scpr2}) vanishes for vectors
of the form (\ref{can2}) with $p>h\,$ and either $m\ge h\,$ or
$m\le p-1-h\,.$ Writing similar conditions for the bra vectors we
end up with the following proposition: {\em all null vectors belong
to the set $\,{\cal I}_h\; |0{\cal i}\,$ or $\,{\cal h} 0 | \;{\cal
I}_h\,$ where ${\cal I}_h\,$ is the ideal generated by $[hp]\;,\
[hH]\;,\ q^{hp} + q^{hH}$, and by $h$-th powers of the $a^i_\a\,$
or, equivalently, by the $h$-th powers of ${\tilde a}^\a_i\,$.} The
factor algebra ${\cal A}_h = {\cal A} / {\cal I}_h\,$ is spanned by
monomials of the type
\be
\lb{fact2}
q^{\mu p} q^{\nu H}(a^1_1)^{m_1} (a^1_2)^{m_2} (a^2_1)^{n_1}
(a^2_2)^{n_2}\;,\ -h<\mu\le h\;,\ 0\le\nu <h\;,\ 0\le m_i ,\; n_i
<h
\ee
and is, hence, (not more than) $2h^6\,$ dimensional.

The definition of the ideal ${\cal I}_h\,$ can be generalized for
any $n\ \ge 2 \,$ assuming that it includes the $h$-th powers of
{\em all minors of the quantum matrix} $( a^i_\a )\,$ (for $n=3\,,$
equivalently, the $h$-th powers of $a^i_\a \,$ {\em and} ${\tilde
a}^\a_i \;$). It follows from Eq.(\ref{genex}), taking into account
the vanishing of $[h]\;,\,$ and from (\ref{aa1}) that
\be
\lb{susy1}
(a^i_\a )^h a^j_\b + (-1)^{
\d_{\a\b}} a^j_\b (a^i_\a )^h\, =\, 0\
\ (\, =\, [ \;[p_{ij}]\; , (a^i_\a )^h ]_+ \, )
\ee
implying also
\be
\lb{susy2}
(a^i_\a )^h {\tilde a}^\b_j + (-1)^{\d_\a^\b}
{\tilde a}^\b_j (a^i_\a )^h\ \, = \, 0 \quad for\quad n=3\,.
\ee
Similar relations are obtained (by transposition of (\ref{susy1}),
(\ref{susy2})) for $({\tilde a}^\a_i )^h\,$ thus proving that the
ideal ${\cal I}_h\,$ is indeed nontrivial, ${\cal I}_h \ne {\cal
A}\,.$

One can analyze on the basis of Lemma 3.4 the structure of indecomposable
$U_q(sl_3)\,$ modules for, say, $h<p_{13}<3h\,,$ thus extending the result of
Proposition 3.6. For example, as a corollary of (3.31),
for $q\,$ given by (\ref{0.1}) (a $2h$-th root of $1\;$)
a HWV (a LWV) is annihilated by $F_i\ (E_i )\,$ if
$\lambda_i = 0\,{\rm mod}\, h\ ( \lambda_{\bar i} = 0\,{\rm mod}\, h )\,$ where
$\lambda_{\bar 1} = \lambda_2\,,\ \lambda_{\bar 2} = \lambda_1\,$. If, in
particular, both $\lambda_i\,$ are multiples of $h\,,$ then the
corresponding weight vector spans a one-dimensional IR of $U_q(sl_3)\,.$

For $n>2\,,$ however, the subspace ${\cal I}_h\; | 0{\cal
i}\,$ does not exhaust the set of null vectors in ${\cal F}\,.$
Indeed, for $n=3\,$ it follows from (\ref{scsq3}) and from the
non-degeneracy of the highest and the lowest weight eigenvalues of
the Cartan generators that the HWV and the LWV are null vectors for
$p_{13} > h\,$:
\be
\lb{0}
{\cal h} {\cal F} |\lambda_1 \lambda_2 {\cal i}
= 0 = {\cal h} {\cal F} |-\lambda_2 \, -\lambda_1 {\cal i} \quad
for \quad \lambda_1 +\lambda_2 + 1 = p_{13} - 1 \ge h\,.
\ee
(If the conjecture (\ref{scsq}) is satisfied then the HWV and the
LWV for any $n\,$ are null vectors for $p_{1n} \ge h+1\,$.) Since the
representation of highest weight $(\l_1\;,\, \l_2)\,$ is
irreducible for $\l_i \le h-1\,$ (cf. (3.31)), the subspace ${\cal
N}\;\subset\; {\cal F}\,$ of null vectors contains ${\cal F}_p\,$
for $p_{12} = \l_1+1\le h\,,\ p_{23} =\l_2 +1\le h\,,\
p_{13}= p_{12}+p_{23} > h\,:$
\be
\lb{FP}
{\cal P}_{\l_1\l_2} (a^1_\a ; {a^3_\b}') |0 {\cal i} \in {\cal N}
\quad for\quad \l_i\le h-1\,, \ \l_1+\l_2\ge h-1
\ee
for ${\cal P}_{\l_1\l_2} ( \rho_1 a^1_\a ; \rho_2 {a^3_\b}') =
\rho_1^{\l_1} \rho_2^{\l_2}
{\cal P}_{\l_1\l_2} (a^1_\a ; {a^3_\b}')\,,$ -- i.e., for any
homogeneous polynomial ${\cal P}_{\l_1\l_2}\,$ of degree $\l_1\,$
in the first three variables, $a^1_\a\,,$ and of degree $\l_2\ge
h-\l_1 -1\,$ in ${a^3_\b}'\,$.
It follows that ${\cal N}\,$ contains all
$U_q\,$ modules ${\cal F}_{\tilde p}\,$ of weights
(\ref{singvect}) corresponding to the first Kac-Moody singular vector
for $p_{13} < h\ (\; \Rightarrow\ {\tilde p}_{13}
= 2h - p_{13} > h\;$ -- see Remark 2.1). Hence, the factor space
${\cal F} / {\cal N}\,$ would be too small to accommodate the
gauge theory treatment of the zero mode counterpart of such
singular vectors.

We can write the null space ${\cal N}\,$ in the form ${\cal N}
= {\tilde{\cal I}}_h | 0 {\cal i}\,$ where ${\tilde{\cal I}}_h
\subset {\cal A}\,$ is the ideal containing all ${\cal
P}_{\l_1\l_2}\,$ appearing in (\ref{FP}) and
closed under transposition, which contains ${\cal I}_h\,$ as a
proper subideal. (We note that the transposition (\ref{'}) is ill
defined for $q\,$ a root of unity whenever ${\cal D}_i(p)\,$
vanishes.) The above discussion induces us to define the factor
algebra
\be
\lb{fa}
{\cal A}_h = {\cal A} / {\cal I}_h
\ee
(rather than ${\cal A} / {\tilde{\cal I}}_h\,$) as the
restricted zero-mode algebra for $q\,$ a root of unity. It is
easily verified (following the pattern of the $n=2\,$ case) that
${\cal A}_h\,$ is again a finite dimensional algebra. Its Fock
space ${\cal F}^h\,$ includes vectors of the form
\be
\lb{6a}
(a^2_1)^{m_1}\; (a^2_2)^{m_2}\; (a^2_3)^{m_3}\;
(a^1_1)^{n_1}\; (a^1_2)^{n_2}\; (a^1_3)^{n_3}\; | 0 {\cal i}
\ee
for $m_i , n_i < h\ (\sum_i m_i \le \sum_i n_i\; )\,$ thus allowing
for weights
\be
\lb{3h}
p_{13} = n_1 +n_2 + n_3 + 2 \le 3h-1\,.
\ee
This justifies the problem
of studying indecomposable
$U_q(sl_3)\,$ modules for $p_{13} < 3h\,$.

\vspace{5mm}

{\em To sum up,} the intertwining quantum matrix algebra ${\cal
A}\,$ introduced in \cite{HIOPT} is an appropriate tool for
studying the WZNW chiral zero modes. Its Fock space representation
provides the first known model of $U_q\,$ for generic $q\,$.
For exceptional $q\,$ (satisfying (\ref{0.1})) it gives room -- by
the results of this section -- to the "physical $U_q\,$ modules"
coupled to the integrable (height $h\,$) representations of the
${\widehat{su}}(n)\,$ Kac-Moody algebra. This is a prerequisite
for a BRS treatment of the zero
mode problem of the two dimensional WZNW model (carried out, for
$n=2\,,$ in \cite{DT}).

\vspace{5mm}


\section*{\bf Acknowledgements}

\medskip

The final version of this paper has been completed during visits of
O.V.O. at ESI, Vienna, of L.K.H. at ESI, Vienna and at ICTP and
INFN, Trieste, and of I.T.T. at ESI, Vienna, at SISSA, Trieste and
at IHES, Bures-sur-Yvette. The authors thank these institutions for
hospitality.
P.F. acknowledges the support of the Italian Ministry of
Education, University and Research (MIUR).
This work is supported in part by the Bulgarian National Council for
Scientific Research under contract F-828, and by CNRS and RFBR
grants PICS-608, RFBR 98-01-22033. The work of A.P.I. and P.N.P.
is supported in part by RFBR grant No. 00-01-00299.

\vspace{10mm}


\section*{Appendix A.  Monodromy matrix and identification of
$U_q(sl_n)\,$ generators for $n=2\,$ and $n=3\,$}
\setcounter{equation}{0}
\def\theequation{A.\arabic{equation}}

\medskip
Eq.(2.11) rewritten as
\be
\lb{A.1}
M=a^{-1} M_p a\quad {\rm or}\quad M^\a_\b =\sum_{i=1}^n
(a^{-1})^\a_i a^i_\b q^{-2p_i - 1+{1\over n}}\,,
\ee
together with the Gauss decomposition (\ref{1.5}) of the monodromy
allows to express by
(\ref{dEF}) the Chevalley generators of $U_q\,$
as well as the operators $E_i E_{i+1} - q E_{i+1} E_i\,,\ \
F_{i+1} F_i - q F_i F_{i+1}\,$ etc. as
linear combinations of products $a^1_{\a_1}\,\dots
a^n_{\a_n}\,$
(with coefficients that depend on $q^{p_i}\,$ and the Cartan elements
$q^{\pm H_i}\,$).
Indeed, in view of (\ref{qdet}), (\ref{qdnorm}), we can express
the elements of the inverse
quantum matrix in terms of the (noncommutative)
algebraic complement ${\tilde a}^\alpha_i\,$ of $a^i_\alpha\,:$
\be
\lb{A.2}
{\cal D}(p) (a^{-1})^\a_i =
{\tilde a}^\a_i =
{{(-1)^{n-1}}\over{[n-1]!}} \varepsilon_{i_1\dots i_{n-1} i}{\cal E}^{\a
\a_1\dots
\a_{n-1}} a^{i_1}_{\a_1}\dots a^{i_{n-1}}_{\a_{n-1}}\,.
\ee
Eq.(\ref{A.2}) is equivalent to (\ref{'}) since for the constant
${\varepsilon}$-tensor used here we have $(-1)^{n-1}
\varepsilon_{i_1\dots i_{n-1} i} =
\varepsilon_{i i_1\dots i_{n-1}}\,.$
Thus we can recast (2.30-33) and (\ref{A.1})
in the form
\be
\lb{A.3}
{\tilde a}^\a_i a^i_\b =
{\cal D}(p)\d^\a_\b\,,\quad \sum_{i=1}^n {\tilde a}^\a_i a^i_\b
q^{-2p_i -1+{1\over n}} = {\cal D}(p) M^\a_\b\,.
\ee
Using Eqs.(4.10-12) of \cite{HIOPT} we can also write
\be
\lb{A.4}
{1\over{{\cal D}(p)}} a^i_\a {\tilde a}^\a_j = N^i_j (p) =
\d^i_j \prod_{k<i} {{[p_{ki}+1]}\over{[p_{ki}]}}
\prod_{i<l} {{[p_{il}-1]}\over{[p_{il}]}}\,.
\ee
We can express the $U_q\,$ generators in terms of products ${\tilde
a}^\a_i a^i_\b\,$ (no summation over $i\;$).
To this end we use (\ref{1.5}), (\ref{1.6}) to
write
\be
\lb{A.5}
M^\a_\b \, =\,q^{{1\over n}-n}\;
\sum_{\sigma = {\rm max}\; (\a,\b )}^n \, f_{\a\; \sigma -1} d_\sigma
e_{\sigma -1\; \b} d_\b
\ee
with $f_{\a\;\a} = f_\a\,,\ e_{\a\;\a} = e_\a\,;\ f_{\a\;\a -1} =
1 = e_{\a -1\;\a}\ $
(see (\ref{dEF})). It is thus simpler to start the
identification of the elements with $M^n_{\b}\,$ and $M^{\a}_n\,.$
Using (\ref{dEF}), we find, in particular,
$${d_n}^2 = q^{2\Lambda_{n-1}} =
{1\over{{\cal D}(p)}} \sum_{i=1}^n\;{\tilde a}^n_i a^i_n {q^{n-1-2{p}_i}}
\,.
$$
We shall spell out the full set of resulting relations for $n=2\,$
and $n=3\,.$

The general relation between Cartan generators and $sl_n\,$ weights
\be
\lb{A.6}
H_i \equiv \sum_j c_{ij} \Lambda_j = 2\Lambda_i -\Lambda_{i-1} -\Lambda_{i+1}
\quad (\,\Lambda_0 \, = \, \Lambda_n\, = 0\, )
\ee
tells us, for $n=2\,,$ that $2\Lambda_1 = H\,.$ This allows
(using (1.28b) and (\ref{dEF})) to write the relations (\ref{A.3})
in the form
\be
\lb{A.7}
{\tilde a}^\a_i a^i_\b = [p] \d^\a_\b\,,
\ee
\be
\lb{A.8}
\bq^p {\tilde a}^\a_1 a^1_\b + q^p {\tilde a}^\a_2 a^2_\b = \bq [p] (M_+
M^{-1}_- )^\a_\b\,.
\ee
Inserting for $M_+ M_-^{-1}\,$ (\ref{1.5}), (\ref{1.6}) and (1.24) we find
for $n=2\,$
\ba
\lb{M2}
&&\bq M_+ M_-^{-1} = \bq
\left(\matrix{ \bq^{H\over 2}& (\bq -q) F q^{H\over 2}\cr
0 & q^{H\over 2}\cr}\right) \;
\left(\matrix{ \bq^{H\over 2}& 0\cr
(\bq -q) E \bq^{H\over 2}& q^{H\over 2}\cr}\right) =\nonumber\\
&&= \left(\matrix{ q^p+\bq^p-q^{H+1}& (\bq -q) E'\cr
(\bq -q)E & q^{H-1}\cr}\right)\,, \quad E' = F q^{H-1} \,.
\ea
As a result we obtain
\ba
\lb{A.10}
\bq^p {\tilde a}^1_1 a^1_1 + q^p {\tilde a}^1_2 a^2_1
&=& [p ] (q^{p} + \bq^{p} - q^{H+1})\,,\nonumber\\
\bq^p {\tilde a}^2_1 a^1_2  + q^p {\tilde a}^2_2 a^2_2
&=& [p ] q^{H-1}\,,\\
\bq^p {\tilde a}^1_1 a^1_2 + q^p {\tilde a}^1_2 a^2_2
&=& [p ] (\bq - q) E'\,,\nonumber\\
\bq^p {\tilde a}^2_1 a^1_1  + q^p {\tilde a}^2_2 a^2_1
&=& [p ] (\bq - q) E\,.\nonumber
\ea
Together with (\ref{A.7}) this gives $8$ equations for the $8$ products
${\tilde a}^\a_i a^i_\b\,$ which can be solved with the result
\ba
\lb{A.11}
{\tilde a}^1_1 a^1_1 = {{q^{H+1}-{\bq}^{p}}\over{q-\bq}}\,
(= q a^2_2 {\tilde a}^2_2 )\,
&,&\quad
{\tilde a}^1_2 a^2_1 = {{q^{p}-q^{H+1}}\over{q-\bq}}\,
(= q a^1_2 {\tilde a}^2_1 )\,\nonumber\\
{\tilde a}^2_2 a^2_2 = {{q^{H-1}-{\bq}^{p}}\over{q-\bq}}\,
(= \bq a^1_1 {\tilde a}^1_1 )\,&,&\quad
{\tilde a}^2_1 a^1_2 = {{q^{p}-q^{H-1}}\over{q-\bq}}\,
(= \bq a^2_1 {\tilde a}^1_2 )\,;\\
{\tilde a}^2_1 a^1_1 = E = - {\tilde a}^2_2 a^2_1\,
(= a^1_1 {\tilde a}^2_1 )\,&,&\quad
{\tilde a}^1_1 a^1_2 = E' = - {\tilde a}^1_2 a^2_2\,
(= a^1_2 {\tilde a}^1_1 )\,\nonumber
\ea
further implying
\ba
\lb{A.12}
a^2_2 {\tilde a}^2_2 - a^1_1 {\tilde a}^1_1 = &\bq^p& =
\bq {\tilde a}^1_1 a^1_1 - q {\tilde a}^2_2 a^2_2\,,\nonumber\\
a^2_1 {\tilde a}^1_2 - a^1_2 {\tilde a}^2_1 = &q^p& =
q {\tilde a}^2_1 a^1_2 - \bq {\tilde a}^1_2 a^2_1\,,\\
{\tilde a}^1_1 a^1_1 - {\tilde a}^2_2 a^2_2 = &q^H& =
{\tilde a}^2_1 a^1_2 - {\tilde a}^1_2 a^2_1\,.\nonumber
\ea
In deriving the relations including products of the type
$a^i_\a {\tilde a}^\b_i\,$ (appearing in parentheses in (\ref{A.11})), we
have used (\ref{A.2}) and (\ref{aa1}).

These relations agree with (\ref{aa1}), (\ref{aa2})
for ${\tilde a}^\a_i\,$ given by
(\ref{A.2}) which becomes
\be
\lb{A.13}
{\tilde a}^\a_i = {\cal E}^{\a\b} \varepsilon_{ij} a^j_\b\,,\
{\rm i.e.,}\
{\tilde a}^1_1 = q^{1/2} a^2_2\,,\ {\tilde a}^1_2 =
- q^{1/2} a^1_2\,,\
{\tilde a}^2_1 = - {\bq}^{1/2} a^2_1\,,\ {\tilde a}^2_2 =
{\bq}^{1/2} a^1_1\,,
\ee
implying
\ba
\lb{A.14}
{\tilde a}^1_1 a^1_1 = q a^2_2 {\tilde a}^2_2\,,\quad
{\tilde a}^2_2 a^2_2 = \bq a^1_1 {\tilde a}^1_1\,,\nonumber\\ \\
{\tilde a}^2_1 a^1_2 = \bq a^2_1 {\tilde a}^1_2\,,\quad
{\tilde a}^1_2 a^2_1 = q a^1_2 {\tilde a}^2_1\,.\nonumber
\ea

In the case of $n=3\,$ we make (\ref{A.3}) and (\ref{A.4})
explicit by noting the identities
\be
\lb{A.15}
3p_1=p_{12}+p_{13}\,,\quad 3p_2=p_{23}-p_{12}\,,\quad
3p_3=-p_{13}-p_{23}\,,
\ee
\be
\lb{A.16}
{\cal D}(p)={\cal D}(p_1,p_2,p_3)=[p_{12}][p_{23}][p_{13}]\,,
\ee
\vspace{1mm}
\be
\lb{A.17}
{\cal D}(p) N^i_j (p) =
\ee
$$
={\rm diag}\left( [p_{23}][p_{12}-1][p_{13}-1]\;,
[p_{13}][p_{12}+1][p_{23}-1]\;, [p_{12}][p_{13}+1][p_{23}+1]
\right)
$$

\vspace{2mm}
\noindent
$(N^i_i (p) = [3])\,$. We find, in particular,
\be
\lb{A.18}
{\cal D}(p) q^{2\Lambda_2 - 2} = {\tilde a}^3_1 a^1_3 {\bq}^{{2\over 3}(p_{12}+p_{13})}+{\tilde a}^3_2 a^2_3
q^{{2\over 3}(p_{12}-p_{23})}+{\tilde a}^3_3 a^3_3 q^{{2\over
3}(p_{13}+p_{23})} \,,
\ee
$${\cal D}(p) ({\bq}^2-1) q^{\Lambda_1} E_2 =
{\tilde a}^3_1 a^1_2 {\bq}^{{2\over 3}(p_{12}+p_{13})}+
{\tilde a}^3_2 a^2_2 {q}^{{2\over 3}(p_{12}-p_{23})}+
{\tilde a}^3_3 a^3_2 {q}^{{2\over 3}(p_{13}+p_{23})}\,,
$$
etc.
\vspace{5mm}

\section*{Appendix B. Transposition in
${\cal A}\,$ for $n=3$}
\setcounter{equation}{0}
\def\theequation{B.\arabic{equation}}

\medskip

The involutivity of the transposition (\ref{'}) is easily verified
for $n=2\,$. Here we shall verify it for $n=3\,$ which is
indicative for the general case.

\vspace{5mm}

\noindent {\bf Proposition A.1}~{\em The (linear)
antihomomorphism of ${\cal A}\,$ defined by (\ref{'})
is involutive:} ${a^i_\a}'' =a^i_\a\,.$

\vspace{5mm}

\noindent
{\em Proof}~~Starting with the relation (\ref{'}) for
\be
\lb{B.1}
{\tilde a}^\a_i\, =\, {1\over [2]}\; \varepsilon_{ijk}\; {\cal E}^{\a\b\gamma}
\; a^j_\b \; a^k_\gamma
\ee
we shall prove, say, for $i=1\,,$ that
\ba
\lb{B.2}
&&[2]\; [p_{23}]\; {a^1_\a}'' \, = \, {\cal E}_{\a\b\gamma} (a^3_\b
a^2_\gamma - a^2_\b a^3_\gamma )' =\noindent\\
&&= {1\over [2]} {\cal E}_{\a\b\gamma}{\cal E}^{\gamma\rho\sigma}
\{ {1\over [p_{13}]}\; (a^1_\rho a^3_\sigma - a^3_\rho a^1_\sigma )\;
{{{\tilde a}^\b_3}\over {[p_{12}]}} -
{1\over [p_{12}]}\; (a^2_\rho a^1_\sigma - a^1_\rho a^2_\sigma )\;
{{{\tilde a}^\b_2}\over {[p_{13}]}} \}\,.\nonumber
\ea
Noting the relation between the contraction of two Levi-Civita
tensors and the $q$-antisymmetrizer (1.28b),
\be
\lb{B.3}
{\cal E}_{\a\b\gamma} {\cal E}^{\gamma\rho\sigma} =
A^{\rho\sigma}_{\a\b} = {\bq}^{\epsilon_{\a\b}} \d^{\rho\sigma}_{\a\b}
- \d^{\rho\sigma}_{\b\a}\,,
\ee
we can rewrite (\ref{B.2}) as
\ba
\lb{B.4}
&&[2]^2\; {\cal D}(p)\; {a^1_\a}'' =
{{[p_{12}]}\over{[p_{12}-1]}}
\{\; {\bq}^{\epsilon_{\a\b}} (a^1_\a a^3_\b - a^3_\a a^1_\b ) -
a^1_\b a^3_\a + a^3_\b a^1_\a )\;\}
{\tilde a}^\b_3\; +\nonumber\\
&&+ {{[p_{13}]}\over{[p_{13}-1]}}
\{\; {\bq}^{\epsilon_{\a\b}} (a^1_\a a^2_\b - a^2_\a a^1_\b ) -
a^1_\b a^2_\a + a^2_\b a^1_\a )\;\} {\tilde a}^\b_2\,.
\ea
Applying four times Eq.(\ref{aa2}) in the form
\ba
\lb{B.5}
a^1_\b a^i_\a = {{[p_{1i}-1]}\over{[p_{1i}]}} a^i_\a a^1_\b +
{{{\bq}^{\epsilon_{\a\b}}}\over{[p_{1i}]}} a^1_\a a^i_\b\,,\nonumber\\
a^i_\b a^1_\a = {{[p_{1i}+1]}\over{[p_{1i}]}} a^1_\a a^i_\b -
{{{q}^{\epsilon_{\a\b}}}\over{[p_{1i}]}} a^i_\a a^1_\b\,,\quad
\ea
for $i=2,3, \a\ne\b\,,$ and using (\ref{A.4}),
(\ref{A.16}) and the identities
\be
\lb{B.6}
{\bq}^\epsilon [p] + [p+1] - {\bq}^{\epsilon p} = [2][p] =
{\bq}^\epsilon [p] + [p-1] + {q}^{\epsilon p}
\ee
for $\epsilon = \pm 1\,$, we find that (\ref{B.2}) is equivalent
to
\ba
\lb{B.7}
&&[2]{\cal D}(p) {a^1_\a}'' =\nonumber\\
&&= {{[p_{12}]}\over{[p_{12}-1]}} a^1_\a
[p_{12}] [p_{13}+1][p_{23}+1] +
{{[p_{13}]}\over{[p_{13}-1]}} a^1_\a
[p_{13}] [p_{12}+1][p_{23}-1]=\nonumber\\
&&= [p_{12}][p_{13}]
(\; [p_{23}+1] + [p_{23}-1]\; ) \; a^1_\a\, =\,
[2]{\cal D}(p) a^1_\a \,.
\ea
The last equality is satisfied due to the CR (\ref{qa})
and the "$q$-formula"
$$[p-1]+[p+1]=[2][p]\,.$$

\vspace{5mm}

\section*{Appendix C.  Computation of the scalar square of highest and
lowest weight vectors in the $n=3\,$ case}
\setcounter{equation}{0}
\def\theequation{C.\arabic{equation}}

\medskip

According to the general definition (\ref{monom}), the scalar
square of the HWV in the $U_q(sl_3)\,$ module ${\cal F}_p\,,$
$${\cal h} HWV(p) | HWV (p) {\cal i} =
{\cal h}\lambda_1 \lambda_2 | \lambda_1 \lambda_2 {\cal i}
\quad ( p_{12} = \lambda_1 + 1\,,\ p_{23} = \lambda_2 +1)$$
is given by
\ba
\lb{C.1}
&&{\cal h}\lambda_1 \lambda_2 | \lambda_1 \lambda_2 {\cal i} =
{\cal h} 0 |
(q a^3_3 )^{\lambda_2}
({a^1_1}')^{\lambda_1}
({a^1_1})^{\lambda_1}
(q {a^3_3}' )^{\lambda_2}
| 0 {\cal i} =\nonumber\\
&&= q^{2\lambda_2} {\cal h} 0 |
({a^1_1}')^{\lambda_1}
(a^3_3 )^{\lambda_2}
({a^3_3}' )^{\lambda_2}
({a^1_1})^{\lambda_1}
| 0 {\cal i}
\ea
where
\be
\lb{C.2}
q [p_{12}+1] {a^3_3}' = \bq^{1\over 2} a^2_2 a^1_1 -
q^{1\over 2} a^2_1 a^1_2\,,
\ee
\be
\lb{C.3}
\bq [p_{23}+1] {a^1_1}' = \bq^{1\over 2} a^3_3 a^2_2  -
q^{1\over 2} a^3_2 a^2_3\,.
\ee
We shall prove (\ref{scsq3}) in four steps.

\vspace{5mm}

\noindent
{\bf Step 1}~{\em The exchange relation
\be
\lb{C.4}
a^3_3 {a^3_3}' = {{[p_{23}][p_{13}]}\over{[p_{23}-1][p_{13}-1]}}
{a^3_3}' a^3_3 + B_1 a^3_1 + B_2 a^3_2
\ee
where
\ba
\lb{C.5}
&&q^{3\over 2} [p_{12}+1] B_1 = \bq^{p_{23}}[p_{13}+1] a^2_3 a^1_2 -
\bq^{p_{13}}a^2_2 a^1_3\,,\nonumber\\ \\
&&q^{1\over 2} [p_{12}+1] B_2 = \bq^{p_{13}}a^2_1 a^1_3
-\bq^{p_{23}}[p_{13}+1] a^2_3 a^1_1\,,
\nonumber
\ea
obtained by repeated application of (\ref{aa2}), implies}
\be
\lb{C.6}
a^3_3 ({a^3_3}')^{\lambda_2} (a^1_1 )^{\lambda_1} | 0{\cal i} =
{{[\lambda_2 ][\lambda_1 +\lambda_2 +1]}\over{[\lambda_1 +2]}}
({a^3_3}')^{\lambda_2 -1} a^3_3 (a^1_1)^{\lambda_1} {a^3_3}'
|0{\cal i}\,.
\ee

\vspace{5mm}

\noindent
{\em Proof}~~The last two terms in (\ref{C.4}), proportional to
$a^3_1\,$ and $a^3_2\,,$ do not contribute to (\ref{C.6}) since,
when moved to the right, they yield expressions proportional to
$a^3_\a {a^3_3}' |0{\cal i}\ ( = 0\ {\rm for}\ \a =1,2\,$).
Repeated application of (\ref{C.4}) in which only the first term in
the right hand side is kept gives (\ref{C.6}).

\vspace{5mm}

\noindent
{\bf Step 2}~{\em The exchange relation
\be
\lb{C.7}
[p_{ij}-m] a^j_\a (a^i_\b)^m=[p_{ij}](a^i_\b)^m a^j_\a -
q^{\epsilon_{\b\a}(p_{ij}-m+1)}[m](a^i_\b)^{m-1}a^i_\a a^j_\b
\ee
which is a consequence of (\ref{aa2}), implies}
\be
\lb{C.8}
a^3_3 (a^1_1)^{\lambda_1}{a^3_3}' |0{\cal i} =
{{[p_{13}]}\over{[p_{13}-\lambda_1 ]}}(a^1_1)^{\lambda_1}
a^3_3 {a^3_3}' |0{\cal i} =
\bq^2 [\lambda_1 +2] (a^1_1)^{\lambda_1} |0{\cal i}\,.
\ee

\vspace{5mm}

\noindent
{\em Proof}~~Eq.(\ref{C.7}) is established by induction in $m\,.$
Eq.(\ref{C.8}) then follows from the identity $q^2 a^3_3 {a^3_3}'
|0{\cal i} = [2]|0{\cal i}\,.$

\vspace{5mm}

\noindent
{\bf Step 3}~{\em Applying $\lambda_2\,$ times
steps $1$ and $2$ one gets}
\be
\lb{C.9}
q^{2\lambda_2} (a^3_3)^{\lambda_2} ({a^3_3}')^{\lambda_2}
(a^1_1)^{\lambda_1} |0{\cal i} =
{{[\lambda_2 ]![\lambda_1 + \lambda_2 +1]!}\over{[\lambda_1 +1 ]!}}
(a^1_1)^{\lambda_1} |0{\cal i}\,.
\ee

\vspace{5mm}

\noindent {\bf Step 4}~{\em Eqs. (\ref{genex}), (\ref{C.7}) and
(\ref{aa2}) imply
\be
\lb{C.10}
({a^1_1}') (a^1_1)^{\lambda_1}
|0{\cal i} = [\lambda_1 ][\lambda_1 +1] (a^1_1)^{\lambda_1 -1}
|0{\cal i} \,; \ee as a result,} \be \lb{C.11} {\cal h}\lambda_1
0 | \lambda_1 0 {\cal i} = {\cal h}0| ({a^1_1}')^{\lambda_1}
(a^1_1)^{\lambda_1} |0 {\cal i} = [\lambda_1 ]! [\lambda_1 +1]!\,.
\ee

\vspace{5mm}

The last two steps are obvious.

\vspace{5mm}

An analogous computation gives the same result (\ref{scsq3}) for the
scalar square of the LWV
${\cal h}-\lambda_2 \, -\lambda_1 | -\lambda_2\, -\lambda_1 {\cal i}\,$.

\vspace{5mm}


\end{document}